\documentstyle[12pt]{article}
\begin{document}
\hfill{CCUTH-96-07}\par
\hfill{hep-ph/9701233}\par
\vskip 0.5cm
\centerline{\large{{\bf Factorization theorems, effective field theory,
}}}\par 
\centerline{\large{{\bf and nonleptonic heavy meson decays}}}\par 
\vskip 1.0cm
\centerline{Tsung-Wen Yeh}
\vskip 0.3cm
\centerline{Department of Physics, National Cheng-Kung University,}
\centerline{Tainan, Taiwan, R.O.C.}
\vskip 0.3cm
\centerline{Hsiang-nan Li}
\vskip 0.3cm
\centerline{Department of Physics, National Chung-Cheng University,}
\centerline{Chia-Yi, Taiwan, R.O.C.}
\vskip 1.0cm
\centerline{\today}
\vskip 0.3cm
PACS numbers: 13.25.Hw, 11.10.Hi, 12.38.Bx, 13.25.Ft
\vfill
\centerline{\bf Abstract}
The nonleptonic heavy meson decays $B\to D^{(*)}\pi(\rho), J/\psi K^{(*)}$ 
and $D\to K^{(*)}\pi$ are studied based on the three-scale perturbative QCD 
factorization theorem developed recently. In this formalism the 
Bauer-Stech-Wirbel parameters $a_1$ and $a_2$ are treated as the Wilson 
coefficients, whose evolution from the $W$ boson mass down to the  
characteristic scale of the decay processes is determined by effective field
theory. The evolution from the characteristic scale to a lower hadronic
scale is formulated by the Sudakov resummation. The scale-setting ambiguity,
which exists in the conventional approach to nonleptonic heavy meson decays,
is moderated. Nonfactorizable and nonspectator contributions are taken 
into account as part of the hard decay subamplitudes. Our formalism is 
applicable to both bottom and charm decays, and predictions, including 
those for the ratios $R$ and $R_L$ associated with the $B\to J/\psi K^{(*)}$ 
decays, are consistent with experimental data.

\newpage
\centerline{\bf I. INTRODUCTION}
\vskip 0.3cm

The analysis of exclusive nonleptonic heavy meson decays has been a
challenging subject because of the involved complicated QCD dynamics. 
These decays occur through the Hamiltonian
\begin{eqnarray}
H=\frac{G_F}{\sqrt 2}V_{ij}V_{kl}^{*}({\bar q}_l q_k)
({\bar q}_j q_i)\;,
\label{full}
\end{eqnarray}
with $G_F$ the Fermi coupling constant, $V$'s the Cabibbo-Kabayashi-Maskawa 
(CKM) matrix elements, $q$'s the relevant quarks and $({\bar q} q)=
{\bar q} \gamma_\mu(1-\gamma_5)q$ the $V-A$ current. Hard gluon corrections 
cause an operator mixing, and their renormalization-group (RG) summation 
leads to the effective Hamiltonian
\begin{eqnarray}
H_{\rm eff}=\frac{G_F}{\sqrt 2}V_{ij}V_{kl}^{*}[c_1(\mu)O_1+ 
c_2(\mu)O_2]\;,
\label{eff}
\end{eqnarray}
where the four-fermion operators $O_{1,2}$ are written as 
$O_1=({\bar q}_l q_k)({\bar q}_j q_i)$ and 
$O_2=({\bar q}_j q_k)({\bar q}_l q_i)$. The Wilson coefficients $c_{1,2}$, 
organizing the large logarithms from the hard gluon corrections to all 
orders, describe the evolution from the $W$ boson mass $M_W$ to a lower
scale $\mu$ with the initial conditions $c_1(M_W)=1$ and $c_2(M_W)=0$.

The simplest and most widely adopted approach to exclusive nonleptonic 
heavy meson decays is the Bauer-Stech-Wirbel (BSW) model \cite{BSW} based 
on the factorization hypothesis, in which the decay rates are expressed in 
terms of various hadronic transition form factors. Employing the Fierz 
transformation, the coefficient of the form factors corresponding to the 
external $W$ boson emission is $a_1=c_1+c_2/N$, and that corresponding to 
the internal $W$ boson emission is $a_2=c_2+c_1/N$, $N$ being the number of 
colors. The form factors may be related to each other by heavy quark 
symmetry, and be modelled by different ansatz. The nonfactorizable 
contributions which can not be expressed in terms of hadronic transition 
form factors, and the nonspectator contributions from the $W$ boson exchange
(or annihilation) are neglected. In this way the BSW method avoids 
the complicated QCD dynamics. 

Though the BSW model is simple and gives predictions in fair agreement with 
experimental data, it encounters several difficulties. It has been known 
that the large $N$ limit of $a_{1,2}$, {\it ie.} the choice 
$a_{1}=c_{1}(M_c)\approx 1.26$ and $a_{2}=c_{2}(M_c)\approx -0.52$,
with $M_c$ the $c$ quark mass, explains the data of charm decays \cite{BSW}. 
However, the same large $N$ limit of $a_{1}=c_{1}(M_b)\approx 1.12$ and 
$a_{2}=c_{2}(M_b)\approx -0.26$, $M_b$ being the $b$ quark mass, does not 
apply to the bottom case. Even after including the $c_{1,2}/N$ term so that
$a_1=1.03$ and $a_2=0.11$, the BSW predictions are still insufficient
to match the data. To overcome this difficulty, parameters $\chi$, 
denoting the corrections from the nonfactorizable final-state interactions, 
have been introduced \cite{C}. They lead to the effective coefficients
\begin{equation}
a^{\rm eff}_1=c_1+c_2\left(\frac{1}{N}+\chi_1\right)\;,\;\;\;\;
a^{\rm eff}_2=c_2+c_1\left(\frac{1}{N}+\chi_2\right)\;.
\end{equation}
$\chi$ should be negative for charm decays, canceling the color-suppressed 
term $1/N$, and be positive for bottom decays in order to enhance the 
predictions. Unfortunately, the mechanism responsible for this sign change 
has not been understood completely. Furthermore, in such a framework
theoretical predictions depend sensitively on the choice of the scale $\mu$ 
for the Wilson coefficients: Setting $\mu$ to $2M_b$ or $M_b/2$ gives rise 
to a more than 20\% difference.

Equivalently, one may regard $a_{1,2}$ as free parameters, and 
determine them by data fitting. However, the behavior of the transition 
form factors involved in nonleptonic heavy meson decays requires
an ansatz \cite{CT} as stated above, such that the extraction of $a_{1,2}$ 
from experimental data becomes model-dependent. On the other hand, it was 
found that the ratio $a_2/a_1$ from an individual fit to the CLEO data of 
$B\to D^{(*)}\pi(\rho)$ \cite{A} varies significantly \cite{CT}. It was 
also shown that an allowed domain $(a_1,a_2)$ exists for the three classes 
of decays ${\bar B}^0\to D^{(*)+}$, ${\bar B}^0\to D^{(*)0}$ and 
$B^-\to D^{(*)0}$, only when the experimental errors are expanded to a 
large extent \cite{GKKP}. 

Moreover, it has been very difficult to explain the two ratios associated 
with the $B\to J/\psi K^{(*)}$ decays \cite{GKP},
\begin{equation}
R=\frac{{\cal B}(B\to J/\psi K^*)}{{\cal B}(B\to J/\psi K)}\;,\;\;\;\;
R_L=\frac{{\cal B}(B\to J/\psi K_L^*)}{{\cal B}(B\to J/\psi K^*)}\;,
\label{rrl}
\end{equation}
simultaneously in the BSW framework, where ${\cal B}(B\to J/\psi K)$ is
the branching ratio of the decay $B\to J/\psi K$. It was argued that the 
inclusion of nonfactorizable contributions is essential for the resolution 
of this controversy \cite{CM}. Such contributions have been analyzed in 
\cite{CM,WYL2} based on the Brodsky-Lepage approach to exclusive processes 
\cite{BL}, in which the full Hamiltonian in Eq.~(\ref{full}) was 
employed. It was found that the nonfactorizable internal $W$-emission 
amplitudes are of the same order as the factorizable ones. However, this 
naive perturbative QCD (PQCD) approach can not account for the destructive 
interference between the external and internal $W$-emission contributions 
in charm decays. This is obvious from the fact that the 
coefficient associated with the internal $W$ emissions is $a_2=1/N$ in
both bottom and charm decays, and thus does not change sign.

Recently, a modified PQCD formalism has been proposed following the series 
of works \cite{ASY,LY1,L1,L2,WYL1,CL}, where the PQCD formalism 
constructed from the full Hamiltonian $H$ was shown to be applicable to the 
$B\to D^{(*)}$ decays \cite{WYL1} in the fast recoil region of final-state 
hadrons \cite{L1}. It was recognized that nonleptonic heavy meson decays 
involve three scales: the $W$ boson mass $M_W$, the typical scale $t$ of 
the decay processes, and the hadronic scale of order $\Lambda_{\rm QCD}$. 
Accordingly, the decay rates are factorized into three convolution factors: 
the ``harder" $W$-emission function, the hard $b$ quark decay subamplitude, 
and the nonperturbative meson wave function, which are characterized by 
$M_W$, $t$ and $\Lambda_{\rm QCD}$, respectively. Radiative corrections 
then produce two types of large logarithms $\ln(M_W/t)$ and 
$\ln(t/\Lambda_{\rm QCD})$. In this three-scale factorization theorem 
$\ln(M_W/t)$ are summed to give the evolution from $M_W$ down to $t$ 
described by the Wilson coefficients $a_{1,2}(t)$, and 
$\ln(t/\Lambda_{\rm QCD})$ are summed into a Sudakov factor \cite{CS}, 
which describes the evolution from $t$ to the lower hadronic scale. The 
former has been derived in effective field theory, and the latter has 
been implemented by the resummation technique \cite{LY1}. 

This modified PQCD formalism is $\mu$-independent, {\it ie.} RG-invariant, 
and thus the scale-setting ambiguity existing in conventional effective 
field theory is moderated \cite{CL}. As the variable $t$ runs to below 
$M_b$ and $M_c$, the constructive and destructive interferences involved 
in bottom and charm decays, respectively, appear naturally. Furthermore, 
not only the factorizable, but the nonfactorizable and nonspectator 
contributions are taken into account and evaluated in a systematic way. 
With the inclusion of the nonfactorizable contributions, we find that 
$a_{1,2}$ restore their original role of the Wilson coefficients, instead 
of being treated as the BSW free parameters. The branching ratios of 
various decay modes $B\to D^{(*)}\pi(\rho)$ and $D\to K^{(*)}\pi$, and the 
ratios $R$ and $R_L$ associated with the $B\to J/\psi K^{(*)}$ decays can 
all be well explained by our formalism.

In Sec. II we derive the three-scale PQCD factorization theorem, 
concentrating on the separation of the contributions characterized by 
different scales. The incorporation of the Sudakov resummation is briefly 
reviewed. In Sec. III the decays $B\to D^{(*)}\pi(\rho)$ are investigated 
to demonstrate the importance of the nonfactorizable contributions. We then
apply the formalism to the decays $D\to K^{(*)}\pi$ in Sec. IV, and
show that the internal $W$-emission amplitude can become sufficiently
negative in charm decays. In Sec. V we compute the decay rates of
$B\to J/\psi K^{(*)}$ and find that the predictions for $R$ and $R_L$ match 
the data simultaneously. Section VI is the conclusion, where possible 
further improvements and applications of our approach are proposed.

\vskip 2.0cm
\centerline{\bf II. THREE-SCALE FACTORIZATION THEOREMS}
\vskip 0.3cm

In this section we construct the modified PQCD formalism, that embodies
both effective field theory and factorization theorems. The motivation
comes from the fact that the Wilson coefficients $c_{1,2}$ of the effective 
Hamiltonian in Eq.~(\ref{eff}) are explicitly $\mu$-dependent. Since 
physical quantities such as the decay rates, which are expressed as the 
products of $c_{1,2}$ with the matrix elements of the four-fermion operators 
$O_{1,2}$, do not depend on $\mu$, the latter should contain a $\mu$ 
dependence to cancel that of the former. However, such a cancellation has 
never been implemented in any previous analysis of nonleptonic heavy meson 
decays. As stated in the Introduction, the BSW method employs the 
factorization hypothesis \cite{BSW}, under which the matrix elements of 
$O_{1,2}$ are factorized into two hadronic matrix elements of the (axial) 
vector currents $({\bar q}q)$. Since the current is conserved, the hadronic 
matrix elements have no anomalous scale dependence, and thus the $\mu$ 
dependence of the Wilson coefficients remains. To remedy this problem, 
$\mu$ should be chosen in such a way that the factorization hypothesis 
gives dominant contributions. However, the hadronic matrix elements involve 
both a short-distance scale associated with the heavy quark and a 
long-distance scale with the mesons. Naively setting $\mu$ to the heavy 
quark mass will lose large logarithms containing the small scale. It is 
then quite natural that theoretical predictions are sensitive to the value 
of $\mu$ \cite{NRXS,LSW}. 

We shall show that the cancellation of the $\mu$ dependence is explicit in 
our formalism. We begin with the idea of the conventional PQCD 
factorization theorem for the $B\to D$ transition form factors, which
describe the amplitude of a $b$ quark decay into a $c$ quark through the 
current operator $(\bar{c}_{L}\gamma_{\mu} b_{L})$. Radiative corrections 
to these form factors are ultraviolet (UV) finite, because the current is 
not renormalized. However, the corrections give rise to infrared (IR) 
divergences at the same time, when the loop gluons are soft or collinear to 
the light partons in the mesons. These IR divergences should be separated 
from the full radiative corrections and grouped into nonperturbative soft 
functions.

The separation of IR divergences in one of the higher-order diagrams is 
demonstrated by Fig.~1(a), where the bubble represents the lowest-order
decay subamplitude of the $B$ meson. This diagram is reexpressed into two 
terms: The first term, with proper eikonal approximation for quark 
propagators, picks up the IR structure of the full diagram. The second term, 
containing an IR subtraction, is finite. The first term, being IR sensitive, 
is absorbed into a meson wave function $\phi(b,\mu)$, if the radiative 
correction is two-particle reducible, or into a soft function $U(b,\mu)$, 
if the radiative correction is two-particle irreducible as shown in 
Fig.~1(a). Here $b$ is the conjugate variable of the transverse momentum
$k_T$ carried by a valence quark of the meson, and thus can be regarded 
as the transverse extent of the meson. It will become clear later that the 
scale $1/b$ serves as an IR cutoff of the associated loop integral. The 
function $U$ corresponds, in some sense, to the nonfactorizable final-state 
interactions in the literature of nonleptonic heavy meson decays \cite{C}. 
The second term, being IR finite, is absorbed into the hard decay 
subamplitude $H(t,\mu)$ as a higher-order correction, where $t$ is its 
typical scale. 

The above factorization procedure is graphically described by Fig.~1(b), 
where the diagrams in the first parentheses contribute to $H$, and that in 
the second parentheses to $\phi$ or $U$. Below we shall neglect $U$, 
because of the pair cancellation between the diagram in Fig.~1(a) and the 
diagram with the right end of the gluon attaching the lower quark line. The 
combination of these two diagrams leads to an integrand proportional to a 
factor $1-e^{i{\bf l}_T\cdot {\bf b}}$, $l_T$ being the transverse loop 
momentum. It is then apparent that $U$ is unimportant, if the main 
contributions to the form factors came from the small $b$ region. It will 
be shown that the Sudakov factor mentioned in the Introduction exhibits a 
strong suppression at large $b$, and thus justifies the neglect of $U$. On 
the other hand, $U$ involves complicated color flows. Hence, we leave its 
discussion to a separate work \cite{LT}. 

Though the full diagrams are UV finite, the IR factorization introduces
UV divergences into $\phi$ and $H$, which have opposite signs. This 
observation hints a RG treatment of the factorization formula derived above. 
Let $\gamma_\phi$ be the anomalous dimension of $\phi$. Then the anomalous 
dimension of $H$ must be $-\gamma_\phi$. We have the RG equations
\begin{equation}
\mu\frac{d}{d\mu}\phi=-\gamma_\phi=-\mu\frac{d}{d\mu}H\;,
\label{rgph}
\end{equation}
whose solutions are given by
\begin{eqnarray}
\phi(b,\mu)&=&\phi(b,1/b)\exp\left[-\int_{1/b}^\mu\frac{d{\bar\mu}}
{\bar\mu}\gamma_\phi(\alpha_s({\bar\mu}))\right]\;,
\label{phi} \\
H(t,\mu)&=&H(t,t)\exp\left[-\int_{\mu}^t\frac{d{\bar\mu}}
{\bar\mu}\gamma_\phi(\alpha_s({\bar\mu}))\right]\;.
\label{h}
\end{eqnarray}
Equation (\ref{phi}) describes the evolution of $\phi$ from the IR cutoff
$1/b$ to an arbitrary scale $\mu$, and Eq.~(\ref{h}) describes the evolution 
of $H$ from $\mu$ to the typical scale $t$. The physics characterized by 
momenta smaller than $1/b$ is absorbed into the initial condition 
$\phi(b,1/b)$, which is of nonperturbative origin. After the RG treatment,
the large logarithms $\ln(t/\mu)$ in $H$ are grouped into the exponent, and 
thus the initial condition $H(t,t)$ can be computed by perturbation theory. 
Combining Eqs.~(\ref{phi}) and (\ref{h}), the factorization formula 
becomes free of the $\mu$ dependence as indicated by
\begin{equation}
H(t,\mu)\phi(b,\mu)=H(t,t)\phi(b,1/b)
\exp\left[-\int_{1/b}^t\frac{d{\bar\mu}}
{\bar\mu}\gamma_\phi(\alpha_s({\bar\mu}))\right]\;.
\label{phih}
\end{equation}

The effective Hamiltonian $H_{\rm eff}$ in Eq.~(\ref{eff}) can be 
constructed in a similar way. Consider the nonleptonic $b$ quark decays 
through a $W$ boson emission up to $O(\alpha_{s})$. We reexpress the full 
diagram, which does not possess UV divergences because of the current 
conservation and the presence of the $W$ boson propagator, into two terms 
as shown in Fig.~2(a). The first term, obtained by shrinking the $W$ boson 
line into a point, corresponds to the local four-fermion operators 
$O_{1,2}$ appearing in $H_{\rm eff}$, and is absorbed into the hard decay 
subamplitude $H(t,\mu)$. This subamplitude is characterized by momenta 
smaller than the $W$ boson mass $M_W$, that is, by the typical scale $t$ of 
the heavy meson decays, since gluons in $H$ do not ''see" the $W$ boson. The 
second term, characterized by momenta of order $M_W$ due to the subtraction 
term, is absorbed into a ''harder" function $H_r(M_W,\mu)$ (not a 
amplitude). Note that the factorization in $H$ is not complete yet, because 
it still contains IR divergences, {\it ie.} the contributions characterized 
by the hadronic scale.

We then obtain the $O(\alpha_s)$ factorization formula shown in Fig.~2(b), 
where the diagrams in the first parentheses contribute to $H_r$, and those 
in the second parentheses to $H$. This formula should be interpreted as a 
matrix relation because of the mixing between the operators $O_1$ and $O_2$, 
or equivalently, the four-fermion vertex should be regarded as the linearly
combined operators $O_1\pm O_2$, which evolve independently. The 
four-fermion vertex in the denominator means that $H_r$ does not carry Dirac 
and color matrix structures. Similarly, UV divergences are introduced 
in the above factorization procedure, when the $W$ boson line is shrunk, 
and thus both $H$ and $H_r$ need renormalization. The RG improved 
factorizaton formula is written as
\begin{equation}
H_r(M_W,\mu)H(t,\mu)=H_r(M_W,M_W)H(t,t)
\exp\left[\int_{t}^{M_W}\frac{d{\bar\mu}}
{\bar\mu}\gamma_{H_r}(\alpha_s({\bar\mu}))\right]\;,
\label{hrh}
\end{equation}
with $\gamma_{H_r}$ the anomalous dimension of $H_r$. It is easy to 
identify the exponential in Eq.~(\ref{hrh}) as the Wilson coefficient, 
implying that the scale $\mu$ in the Wilson coefficient should be set to the 
hard scale $t$. The function $H_r(M_W,M_W)$ can now be safely taken as its 
lowest-order expression $H^{(0)}_r=1$, since the large logarithms 
$\ln(M_W/\mu)$ have been organized into the exponent. Note that the 
appropriate active flavor number should be substituted into 
$\alpha_s({\bar \mu})$, when $\bar\mu$ evolves from $M_W$ down to $t$. The 
continuity conditions for the transition of $\alpha_s({\bar\mu})$ between 
regions with different active flavor numbers \cite{Buras} are understood.

We are now ready to construct the three-scale factorization theorem by 
combining Eqs.~(\ref{phih}) and (\ref{hrh}). Start with the nonleptonic 
heavy meson decay amplitude up to $O(\alpha_s)$ without integrating out the 
$W$ boson. The IR sensitive functions are first factorized according to 
Fig.~3(a), such that the diagrams in the first parentheses are characterized 
by momenta larger than the IR cutoff. Employing Fig.~2(b) to separate $H_r$, 
we arrive at the factorization formula described by Fig.~3(b). The diagrams 
in the last parentheses are identified as the hard decay subamplitude $H$. 
It is obvious that its anomalous dimension is given by 
$\gamma_H=-\gamma_\phi-\gamma_{H_r}$. Applying the RG analysis to each 
convolution factor, we derive
\begin{equation}
H_r(M_W,\mu)H(t,\mu)\phi(b,\mu)=c(t)H(t,t)\phi(b,1/b)
\exp\left[-\int_{1/b}^t\frac{d{\bar\mu}}
{\bar\mu}\gamma_\phi(\alpha_s({\bar\mu}))\right],
\label{main}
\end{equation}
where the Wilson coefficient $c(t)$ represents the exponential in 
Eq.~(\ref{hrh}). The cancellation of the $\mu$ dependences among the three 
convolution factors is explicit. The two-stage evolutions from $1/b$ to $t$ 
and from $t$ to $M_W$ have been established. We emphasize that the 
Wilson coefficient appears as a convolution factor of the three-scale
factorization formula, which is, however, a constant coefficient (once
its argument $\mu$ is set to a value) in the conventional approach of
effective field theory.

In the leading logarithmic approximation $c_{1,2}$ are given, in terms of 
the combination $c_{\pm}(\mu) = c_{1}(\mu) \pm c_{2}(\mu)$, by
\begin{equation}
c_{\pm}(\mu) = \left[ \frac{\alpha_{s}(M_{W})}{\alpha_{s}(\mu)} \right]
^{\frac{-6\gamma_{\pm}}{33-2n_{f}}} \;,
\end{equation}
with the constants $2\gamma_+=-\gamma_-=-2$, and $n_f$ the number of active 
quark flavors. Below we shall employ the more complicated two-loop 
expressions of $c_{1,2}$ presented in the Appendix A, that include 
next-to-leading logarithms \cite{Buras}.

At last, we explain how to incorporate the Sudakov factor into the above 
factorization formula. The RG solution in Eq,~(\ref{phi}) sums only the 
single logarithms contained in the meson wave function $\phi$. In fact, 
there exist also double logarithms coming from the overlap of collinear and 
soft divergences. Hence, an extra large scale $P$, the meson momentum, 
should be added into $\phi$ as an argument. The scale $P$ is associated
with the collinear divergences, while the small scale $1/b$ is associated 
with the soft divergences as stated at the beginning of this section. 
Before reaching Eq.~(\ref{rgph}), one performs the resummation for these 
double logarithms, and obtain
\begin{equation}
\phi(P,b,\mu)=\phi(b,\mu)\exp[-s(P,b)]\;.
\label{sdc}
\end{equation}
$e^{-s}$ is the Sudakov factor, which exhibits a strong suppression in the
large $b$ region. The single-scale wave function $\phi(b,\mu)$ discussed 
above is then identified as the initial condition of the resummation for the 
two-scale wave function $\phi(P,b,\mu)$. For the detailed derivation of 
Eq.~(\ref{sdc}), refer to \cite{LY1,L1}.

In summary, the large logarithms $\ln(M_W/t)$ are grouped into the Wilson 
coefficients $c_{1,2}$, and $\ln(tb)$ are organized by the resummation 
technique and by the RG method. Combining Eqs.~(\ref{main}) and (\ref{sdc}),
we derive the final expression of the three-scale factorization formula
\begin{eqnarray}
H_r(M_W,\mu)H(t,\mu)\phi(x,P,b,\mu)&=&c(t)H(t,t)\phi(x,b,1/b)
\nonumber \\
& &\times\exp\left[-s(P,b)-\int_{1/b}^t\frac{d{\bar\mu}}
{\bar\mu}\gamma_\phi(\alpha_s({\bar\mu}))\right],
\nonumber\\ 
& &
\label{for}
\end{eqnarray}
where the momentum fraction $x$ associated with a valence quark of the meson
has been inserted.
\vskip 2.0cm

\centerline{\bf III. The $B\to D^{(*)}\pi{(\rho)}$ Decays}
\vskip 0.3cm

In the conventional BSW approach the branching ratios of the exclusive
nonleptonic heavy meson decays are parametrized only by the factorizable 
contributions from the external and internal $W$ emissions as stated in the 
Introduction. The associated nonfactorizable contributions, which can not 
be expressed in terms of hadronic form factors, are ignored. The 
nonspectator contributions from the $W$-exchange (or annihilation) diagrams, 
which may be factorizable or nonfactorizable, are not included either. 
However, the naive PQCD analysis based on the full Hamiltonian in 
Eq.~(\ref{full}) has shown that the nonfactorizable contributions are 
comparable to the factorizable ones for the internal $W$ emissions and for 
the $W$ exchanges \cite{WYL2}. 

In this section we shall investigate the importance of the nonfactorizable 
and nonspectator contributions to exclusive nonleptonic heavy meson decays
employing the more sophiscated three-scale PQCD factorization theorem 
developed in the previous section \cite{CL}.  We evaluate the branching 
ratios of the $B\to D^{(*)}\pi{(\rho)}$ decays, taking into account the 
factorizable, nonfactorizable and nonspectator contributions, and letting 
the Wilson coefficients $c_{1,2}$ evolve according to effective field 
theory. In this framework the external $W$ emissions also give 
nonfactorizable contributions. The relevant effective Hamiltonian is given
by
\begin{eqnarray}
H_{\rm eff}=\frac{G_F}{\sqrt 2}V_{cb}V_{ud}^{*}[c_1(\mu)O_1+ 
c_2(\mu)O_2]\;,
\label{eff1}
\end{eqnarray}
with the four-fermion operators $O_1=({\bar d}u)({\bar c}b)$ and 
$O_2=({\bar c}u)({\bar d}b)$. The full Hamiltonian $H$ is then a special 
case with the choice $c_1=1$ and $c_2=0$. 

We first study the $B\to D^{(*)}\pi$ decays. The analysis of the 
$B\to D^{(*)}\rho$ decays is similar. The factorizable external 
$W$-emission amplitudes define the $B\to D^{(*)}$ transition form factors
$\xi$ through the hadronic matrix elements,
\begin{eqnarray}
\langle D (P_2)|V^\mu|B(P_1)\rangle
&=&\sqrt{M_BM_D}[\xi_+(\eta)(v_1+v_2)^\mu+
\xi_-(\eta)(v_1-v_2)^\mu]\;,
\nonumber \\
\langle D^* (P_2)|V^\mu|B(P_1)\rangle
&=&i\sqrt{M_BM_{D^*}}\xi_V(\eta)
\epsilon^{\mu\nu\alpha\beta}\epsilon^*_\nu v_{2\alpha}v_{1\beta}\;,
\nonumber \\
\langle D^* (P_2)|A^\mu|B(P_1)\rangle&=&\sqrt{M_BM_{D^*}}
[\xi_{A_1}(\eta)(\eta+1)\epsilon^{*\mu}
-\xi_{A_2}(\eta)\epsilon^*\cdot v_1 v_1^\mu
\nonumber \\
& &-\xi_{A_3}(\eta)\epsilon^*\cdot v_1 v_2^\mu]\;.
\label{iwm}
\end{eqnarray}
$P_1$ $(P_2)$, $M_B$ ($M_{D^{(*)}}$) and $v_1$ $(v_2)$ are the momentum, 
the mass, and the velocity of the $B$ $(D^{(*)})$ meson, satisfying the 
relation $P_1=M_Bv_1$ $(P_2=M_{D^{(*)}}v_2)$. $\epsilon^*$ is the 
polarization vector of the $D^*$ meson. The velocity transfer $v_1\cdot v_2$ 
in two-body nonleptonic decays takes the maximal value $\eta=(1+r^2)/(2r)$ 
with $r=M_{D^{(*)}}/M_B$. In the rest frame of the $B$ meson $P_1$ and $P_2$ 
are expressed as $P_1=(M_B/\sqrt{2})(1,1,{\bf 0}_T)$ and 
$P_2=(M_B/\sqrt{2})(1,r^2,{\bf 0}_T)$ \cite{L1}. For the analysis below,
we define $k_1$ ($k_2$) the momentum of the light valence quark in the $B$ 
($D^{(*)}$) meson. $k_1$ may have a minus component $k_1^-$, giving the 
momentum fraction $x_1=k_1^-/P_1^-$, and small transverse components 
${\bf k}_{1T}$. $k_2$ may have a large plus component $k_2^+$, giving 
$x_2=k_2^+/P_2^+$, and small ${\bf k}_{2T}$. The pion then carries the 
momentum $P_3=P_1-P_2$, whose nonvanishing component is only $P_3^-$. One 
of its valence quark carries the fractional momentum $x_3P_3$, and small 
transverse momenta ${\bf k}_{3T}$. In the infinite mass limit of $M_B$ and 
$M_{D^{(*)}}$ the form factors $\xi_i$ with $i=+$, $-$, $V$, $A_1$, 
$A_2$, and $A_3$ obey the relations
\begin{equation}
\xi_+=\xi_V=\xi_{A_1}=\xi_{A_3}=\xi,\;\;\;\;  \xi_-=\xi_{A_2}=0.
\label{iwr}
\end{equation}
$\xi$ is the so-called Isgur-Wise (IW) function \cite{IW}, which is 
normalized to unity at zero recoil $\eta\to 1$ by heavy quark symmetry. 

$\xi_i$ include the contributions from the hadronic matrix element of $O_1$ 
shown in Fig.~4(a) and from the color-suppressed matrix element of $O_2$ 
in Fig.~4(b). Therefore, their factorization formulas involve the Wilson 
coefficient $a_1=c_1+c_2/N$. We define the form factors 
$\xi^{(*)}_{\rm int}$ for the internal $W$-emission diagrams, which include 
the factorizable contributions from the matrix elements of $O_2$ in 
Fig.~4(c), and from the color-suppressed matrix element of $O_1$ in 
Fig.~4(d). These form factors then contain the Wilson coefficient 
$a_2=c_2+c_1/N$. Similarly, we define the form factors $\xi^{(*)}_{\rm exc}$ 
for the $W$-exchange diagrams, which include the factorizable contributions 
from the matrix elements of $O_2$ in Fig.~4(e), and from the 
color-suppressed matrix element of $O_1$ in Fig.~4(f). Hence, these form 
factors also contain the Wilson coefficient $a_2$. 

For the nonfactorizable contributions to the $B\to D^{(*)}\pi$ decays, the 
possible diagrams are exhibited in Fig.~5, which correspond to those in 
Fig.~4. Figs.~5(a), 5(c), and 5(e) do not contribute at $O(\alpha_s)$ simply 
because a trace of odd number of color matrices vanishes. Hence, all the
nonfactorizable contributions come from Figs.~5(b), 5(d), and 5(f), denoted 
by the amplitudes ${\cal M}_b^{(*)}$, ${\cal M}_d^{(*)}$, and 
${\cal M}_f^{(*)}$, respectively, and are thus color-suppressed. Their 
expressions are more complicated, and can not be written in terms of 
hadronic form factors. The amplitudes ${\cal M}^{(*)}_b$ for the 
nonfactorizable external $W$ emissions depend on the Wilson coefficient 
$c_2/N$. They have the same expressions for the charged and neutral $B$ 
meson decays, because replacing the spectator ${\bar u}$ quark in the $B^-$ 
meson by the ${\bar d}$ quark does not change the Feynman rules. The 
amplitudes ${\cal M}^{(*)}_d$ for the nonfactorizable internal $W$ emissions 
contain the Wilson coefficient $c_1/N$. ${\cal M}^{(*)}_f$ for the 
nonfactorizable $W$ exchanges involve the Wilson coefficient $c_1/N$. 

The decay rates of $B\to D^{(*)}\pi$ have the expression
\begin{equation}
\Gamma_i=\frac{1}{128\pi}G_F^2|V_{cb}|^2|V_{ud}|^2M_B^3\frac{(1-r^2)^3}{r}
|{\cal M}_i|^2\;,
\label{dr}
\end{equation}
where $i=1$, 2, 3 and 4 denote the modes $B^-\to D^0\pi^-$, 
${\bar B}^0\to D^+\pi^-$, $B^-\to D^{*0}\pi^-$ and
${\bar B}^0\to D^{*+}\pi^-$, respectively. 
With the above form factors and the nonfactorizable amplitudes, the decay 
amplitudes ${\cal M}_i$ are written as
\begin{eqnarray}
{\cal M}_1&=&f_\pi[(1+r)\xi_+-(1-r)\xi_-]+
f_D\xi_{\rm int}+{\cal M}_b+{\cal M}_d\;,
\label{M1}\\
{\cal M}_2&=&f_\pi[(1+r)\xi_+-(1-r)\xi_-]+f_B
\xi_{\rm exc}+{\cal M}_b+{\cal M}_f\;,
\label{M2}\\
{\cal M}_3&=&\frac{1+r}{2r}f_\pi[(1+r)\xi_{A_1}-(1-r)(r\xi_{A_2}
+\xi_{A_3})]
\nonumber \\
& &+f_{D^*}\xi^*_{\rm int}+{\cal M}^*_b+{\cal M}^{*}_d\;,
\label{M3}\\
{\cal M}_4&=&\frac{1+r}{2r}f_\pi[(1+r)\xi_{A_1}-(1-r)(r\xi_{A_2}
+\xi_{A_3})]
\nonumber \\
& &+f_B\xi^*_{\rm exc}+{\cal M}^*_b+{\cal M}^{*}_f\;,
\label{M4}
\end{eqnarray}
where $f_B$, $f_{D^{(*)}}$, and $f_\pi$ are the $B$ meson, $D^{(*)}$ meson,
and pion decay constants. We have made explicit that the charged $B$ meson 
decays $B^-\to D^{(*)0}\pi^-$ contain the external and internal $W$-emission 
contributions, and the neutral $B$ meson decays ${\bar B}^0\to 
D^{(*)+}\pi^-$ contain the external $W$-emission and $W$-exchange 
contributions.

In the considered maximal recoil region with $P_2^+\gg M_{D^{(*)}}/
\sqrt{2}\gg P_2^-$, we regard the $D^{(*)}$ meson as a light meson
for simplicity \cite{L1}. Double logarithms contained in the $B$ meson, 
$D^{(*)}$ meson and pion wave functions $\phi_B$, $\phi_{D^{(*)}}$, and 
$\phi_\pi$, respectively, are organized into the corresponding Sudakov 
factors using the resummation technique \cite{L1,CS,LS}:
\begin{eqnarray}
\phi_B(x_1,P_1,b_1,\mu)&=&\phi_B(x_1,b_1,1/b_1)\exp[-S_B(\mu)]\;,
\nonumber \\
\phi_{D^{(*)}}(x_2,P_2,b_2,\mu)&=&\phi_{D^{(*)}}(x_2,b_2,1/b_2)
\exp[-S_{D^{(*)}}(\mu)]\;,
\nonumber \\
\phi_\pi(x_3,P_3,b_3,\mu)&=&\phi_\pi(x_3,b_3,1/b_3)\exp[-S_\pi(\mu)]\;,
\label{wp}
\end{eqnarray}
with the exponents
\begin{eqnarray}
S_B(\mu)&=&s(x_1P_1^-,b_1)+2\int_{1/b_1}^{\mu}\frac{d{\bar \mu}}{\bar \mu}
\gamma(\alpha_s({\bar \mu}))\;,
\nonumber \\
S_{D^{(*)}}(\mu)&=&s(x_2P_2^+,b_2)+s((1-x_2)P_2^+,b_2)+2\int_{1/b_2}^{\mu}
\frac{d{\bar \mu}}{\bar \mu}\gamma(\alpha_s({\bar \mu}))\;,
\nonumber \\
S_\pi(\mu)&=&s(x_3P_3^-,b_3)+s((1-x_3)P_3^-,b_3)+
2\int_{1/b_3}^{\mu}\frac{d{\bar \mu}}{\bar \mu}
\gamma(\alpha_s({\bar \mu}))\;.
\label{wpe}
\end{eqnarray}
The quark anomalous dimension $\gamma=-\alpha_s/\pi$, is related to 
$\gamma_\phi=2\gamma$ introduced in Sec. II. The spatial extents $b_i$ of 
the mesons are the Fourier conjugate variables of $k_{iT}$. We shall neglect 
the intrinsic $b$ dependence of the wave functions, denoted by the argument 
$b$, and the $O(\alpha_s(1/b))$ corrections, denoted by the argument $1/b$. 
That is, we assume $\phi(x,b,1/b)=\phi(x)$. The wave functions $\phi_i(x)$, 
$i=B$, $D$, $D^*$, and $\pi$, satisfy the normalization 
\begin{equation}
\int_0^1\phi_i(x)dx=\frac{f_i}{2\sqrt{6}}\;. 
\end{equation}

The exponent $s$ is written as \cite{BS}
\begin{equation}
s(Q,b)=\int_{1/b}^{Q}\frac{d \mu}{\mu}\left[\ln\left(\frac{Q}{\mu}
\right)A(\alpha_s(\mu))+B(\alpha_s(\mu))\right]\;,
\label{fsl}
\end{equation}
where the anomalous dimensions $A$ to two loops and $B$ to one loop are 
given by
\begin{eqnarray}
A&=&{\cal C}_F\frac{\alpha_s}{\pi}+\left[\frac{67}{9}-\frac{\pi^2}{3}
-\frac{10}{27}n_f+\frac{8}{3}\beta_1\ln\left(\frac{e^{\gamma_E}}{2}\right)
\right]\left(\frac{\alpha_s}{\pi}\right)^2\;,
\nonumber \\
B&=&\frac{2}{3}\frac{\alpha_s}{\pi}\ln\left(\frac{e^{2\gamma_E-1}}
{2}\right)\;,
\end{eqnarray}
with ${\cal C}_F=4/3$ the color factor and $\gamma_E$ the Euler constant. 
The two-loop expression of the running coupling constant, 
\begin{equation}
\frac{\alpha_s(\mu)}{\pi}=\frac{4}{\beta_0\ln(\mu^2/\Lambda^2)}-
\frac{16\beta_1}{\beta_0^3}\frac{\ln\ln(\mu^2/\Lambda^2)}
{\ln^2(\mu^2/\Lambda^2)}\;,
\end{equation}
will be substituted into Eq.~(\ref{fsl}), with the coefficients 
\begin{eqnarray}
& &\beta_{0}=\frac{33-2n_{f}}{3}\;,\;\;\;\beta_{1}=\frac{153-19n_{f}}{6}\;,
\label{12}
\end{eqnarray}
and the QCD scale $\Lambda\equiv \Lambda_{\rm QCD}$. 

Combined with the evolution of the hard decay subamplitudes $H$, the 
variables $\mu$ in Eq.~(\ref{wpe}) are replaced by the hard scales $t$ as 
shown in Eq.~(\ref{for}), leading to the RG invariant Sudakov exponents 
$S_B(t)$, $S_{D^{(*)}}(t)$ and $S_\pi(t)$. Since large logarithms have been 
organized, we compute $H$ to lowest order by sandwiching Figs.~4 and 5 with 
the matrix structures $(\not P_1+M_B)\gamma_5/\sqrt{2N}$ from the initial 
$B$ meson, with $\gamma_5(\not P_2+M_D)/\sqrt{2N}$, 
$\not\epsilon(\not P_2+M_{D^*})/\sqrt{2N}$, and 
$\gamma_5\not P_3/\sqrt{2N}$ from the final $D$ meson, $D^{*}$ meson, and 
pion, respectively.

The expressions of all the form factors and nonfactorizable amplitudes for 
the $B\to D^{(*)}\pi$ decays are listed below. The form factors $\xi_i$, 
$i=+$, $A_1$ and $A_3$, and $\xi_j$, $j=-$ and $A_2$, derived from 
Figs.~4(a) and (b), are given by 
\begin{eqnarray}
\xi_i&=& 16\pi{\cal C}_F\sqrt{r}M_B^2
\int_{0}^{1}d x_{1}d x_{2}\,\int_{0}^{\infty} b_1d b_1 b_2d b_2\,
\phi_B(x_1)\phi_{D^{(*)}}(x_2)a_1(t)
\nonumber \\
& &\times \alpha_s(t)[(1+\zeta_ix_2r)h(x_1,x_2,b_1,b_2,m)
+(r+\zeta'_ix_1)h(x_2,x_1,b_2,b_1)]
\nonumber \\
& &\times \exp[-S_B(t)-S_{D^{(*)}}(t)]\;,
\label{+}\\
\xi_j&=& 16\pi{\cal C}_F\sqrt{r}M_B^2
\int_{0}^{1}d x_{1}d x_{2}\,\int_{0}^{\infty} b_1d b_1 b_2d b_2\,
\phi_B(x_1)\phi_{D^{(*)}}(x_2)a_1(t)
\nonumber \\
& &\times \alpha_s(t)[\zeta_j x_2rh(x_1,x_2,b_1,b_2)
+\zeta'_jx_1 h(x_2,x_1,b_2,b_1)]
\nonumber \\
& &\times \exp[-S_B(t)-S_{D^{(*)}}(t)]\;,
\label{-}
\end{eqnarray}
with the constants \cite{WYL1}
\begin{eqnarray}
& &\zeta_+=\zeta'_+=\frac{1}{2}\left[\eta-\frac{3}{2}+
\sqrt{\frac{\eta-1}{\eta+1}}\left(\eta-\frac{1}{2}\right)\right]\;,
\nonumber \\
& &\zeta_-=-\zeta'_-=-\frac{1}{2}\left[\eta-\frac{1}{2}+
\sqrt{\frac{\eta+1}{\eta-1}}\left(\eta-\frac{3}{2}\right)\right]\;,
\nonumber \\
& &\zeta_{A_1}=-\frac{2-\eta-\sqrt{\eta^2-1}}{2(\eta+1)}\;,
\;\;\;\;\zeta'_{A_1}=-\frac{1}{2(\eta+1)}\;,
\nonumber \\
& &\zeta_{A_2}=0\;,\;\;\;\;\zeta'_{A_2}=-1-\frac{\eta}{\sqrt{\eta^2-1}}\;.
\nonumber \\
& &\zeta_{A_3}=-\frac{1}{2}-\frac{\eta-2}{2\sqrt{\eta^2-1}}\;,
\;\;\;\;\zeta'_{A_3}=\frac{1}{2\sqrt{\eta^2-1}}\;,
\end{eqnarray}
Here $\eta$ takes the maximal velocity transfer $\eta=(1+r^2)/(2r)$ as 
stated before. For the behavior of the above form factors at other values of 
velocity transfer, refer to \cite{WYL1}.

The form factors $\xi^{(*)}_{\rm int}$ from Figs.~4(c) and 4(d), and 
$\xi^{(*)}_{\rm exc}$ from Figs.~4(e) and 4(f) are written as
\begin{eqnarray}
\xi^{(*)}_{\rm int}&=&16\pi{\cal C}_F\sqrt{r}M_B^2
\int_0^1 dx_1dx_3\int_0^{\infty}b_1db_1b_3db_3
\phi_B(x_1)\phi_\pi(x_3)a_2(t_{\rm int})
\nonumber \\
& &\times \alpha_s(t_{\rm int})
\left[(1+x_3(1-r^2))h_{\rm int}(x_1,x_3,b_1,b_3,m_{\rm int})
\right.
\nonumber \\
& &\left.+\zeta^{(*)}_{\rm int}x_1r^2 
h_{\rm int}(x_3,x_1,b_3,b_1,m_{\rm int})\right]
\exp[-S_B(t_{\rm int})-S_\pi(t_{\rm int})],
\label{int} \\\
\xi^{(*)}_{\rm exc}&=&16\pi{\cal C}_F\sqrt{r}M_B^2
\int_0^1 dx_2dx_3\int_0^{\infty}b_2db_2b_3db_3
\phi_{D^{(*)}}(x_2)\phi_\pi(x_3)a_2(t_{\rm exc})
\nonumber \\
& &\times\alpha_s(t_{\rm exc})
\left[(x_3(1-r^2)-\zeta^{(*)}_{\rm exc}r^2)
h_{\rm exc}(x_2,x_3,b_2,b_3,m_{\rm exc})\right.
\nonumber \\
& &\left.+x_2h_{\rm exc}(x_3,x_2,b_3,b_2,m_{\rm exc})\right]
\exp[-S_{D^{(*)}}(t_{\rm exc})-S_\pi(t_{\rm exc})]\;,
\label{exc}
\end{eqnarray}
with the constants $\zeta_{\rm int}=\zeta_{\rm exc}=-\zeta^*_{\rm int}
=-\zeta^*_{\rm exc}=1$. In the derivation of 
$\xi_{\rm int}^{(*)}$ we have assumed that $k_1$ has a plus 
component $k_1^+=x_1P_1^+$. It is obvious that $\xi^{(*)}_{\rm int}$ and 
$\xi^{(*)}_{\rm exc}$ are exactly the $B\to\pi$ and $D^{(*)}\to\pi$ 
transition form factors, respectively, evaluated at maximal recoil. 

In Eqs.~(\ref{+}), (\ref{-}), (\ref{int}) and (\ref{exc}) the functions 
$h$'s, obtained from the Fourier transform of the lowest-order $H$, are 
given by
\begin{eqnarray}
h(x_1,x_2,b_1,b_2,m)&=&K_{0}\left(\sqrt{x_1x_2m}b_1\right)
\nonumber \\
& &\times \left[\theta(b_1-b_2)K_0\left(\sqrt{x_2m}
b_1\right)I_0\left(\sqrt{x_2m}b_2\right)\right.
\nonumber \\
& &\left.+\theta(b_2-b_1)K_0\left(\sqrt{x_2m}b_2\right)
I_0\left(\sqrt{x_2m}b_1\right)\right],
\label{dh}\\
h_{\rm int}(x_1,x_3,b_1,b_3,m_{\rm int})&=&h(x_1,x_3,b_1,b_3,
m_{\rm int})\;,
\label{hint}\\
h_{\rm exc}(x_2,x_3,b_2,b_3,m_{\rm exc})&=&\frac{\pi^2}{4}
H_0^{(1)}\left(\sqrt{x_2x_3m_{\rm exc}}b_2\right)
\nonumber \\
& &\times\left[\theta(b_2-b_3)
H_0^{(1)}\left(\sqrt{x_3m_{\rm exc}}b_2\right)
J_0\left(\sqrt{x_3m_{\rm exc}}b_3\right)\right.
\nonumber \\
& &\left.+\theta(b_3-b_2)H_0^{(1)}\left(\sqrt{x_3m_{\rm exc}}b_3\right)
J_0\left(\sqrt{x_3m_{\rm exc}}b_2\right)\right],
\nonumber \\
& &
\end{eqnarray}
with $m=M_B^2$ and $m_{\rm int}=m_{\rm exc}=(1-r^2)M_B^2$. We observe that 
the $W$-exchange contributions are complex due to the exchange of time-like 
hard gluons. The hard scales $t$ are chosen as
\begin{eqnarray}
t&=&{\rm max}(\sqrt{x_1m},\sqrt{x_2m},1/b_1,1/b_2) \\
t_{\rm int}&=&{\rm max}(\sqrt{x_1m_{\rm int}},\sqrt{x_3m_{\rm int}},
1/b_1,1/b_3) \\
t_{\rm exc}&=&{\rm max}(\sqrt{x_2m_{\rm exc}},\sqrt{x_3m_{\rm exc}},
1/b_2,1/b_3)\;. 
\end{eqnarray}
where we consider only the energies $\sqrt{x_im}$ and $\sqrt{x_jm}$ from 
the longitudinal momenta of the internal quarks in the diagrams of Fig.~4, 
because the gluon energies $\sqrt{x_ix_jm}$ are always smaller. The scales 
$1/b_i$ are associated with the transverse momenta. Then Sudakov 
suppression guarantees that the main contributions come from the large $t$ 
region, where the running coupling constant $\alpha_s(t)$ is small, and 
perturbation theory is reliable.

For the nonfactorizable amplitudes, the factorization formulas involve
the kinematic variables of all the three mesons, and the Sudakov exponent 
is given by $S=S_B+S_{D^{(*)}}+S_\pi$. The integration over $b_3$ can be 
performed trivially, leading to $b_3=b_1$ or $b_3=b_2$. Their expressions
are
\begin{eqnarray}
{\cal M}^{(*)}_b&=& 32\pi\sqrt{2N}{\cal C}_F\sqrt{r}M_B^2 
\int_0^1 [dx]\int_0^{\infty}
b_1 db_1 b_2 db_2
\phi_B(x_1)\phi_{D^{(*)}}(x_2)\phi_\pi(x_3) \nonumber \\
& &\times \biggl\{ \alpha_s(t_b^{(1)})\frac{c_2(t_b^{(1)})}{N}  
\exp[-S(t_b^{(1)})|_{b_3=b_2}]
\nonumber \\
& &\hspace{0.5cm}\times
[(1-r^2)(1-x_3)-x_1+\zeta^{(*)}_b(r-r^2)(x_1-x_2)]h^{(1)}_b(x_i,b_i) 
\nonumber \\
& &\hspace{0.5cm}
- \alpha_s(t_b^{(2)})\frac{c_2(t_b^{(2)})}{N}  
\exp[-S(t_b^{(2)})|_{b_3=b_2}]
\nonumber \\
& &\hspace{0.5cm}
\times[(2-r)x_1-(1-r)x_2-(1-r^2)x_3]h^{(2)}_b(x_i,b_i) \biggr\}\;,
\label{mb}\\
{\cal M}^{(*)}_d&=& 32\pi\sqrt{2N}{\cal C}_F\sqrt{r}M_B^2
\int_0^1 [dx]\int_0^{\infty}b_1 db_1 b_2 db_2
\phi_B(x_1)\phi_{D^{(*)}}(x_2)\phi_\pi(x_3) \nonumber \\
& &\times \zeta^{(*)}_d 
\biggl\{ \alpha_s(t_d^{(1)})\frac{c_1(t_d^{(1)})}{N}  
\exp[-S(t_d^{(1)})|_{b_3=b_1}]
\nonumber \\
& &\hspace{0.5cm}
\times[x_1-x_2-x_3(1-r^2)]h^{(1)}_d(x_i,b_i) 
\nonumber \\
& &\hspace{0.5cm}
+ \alpha_s(t_d^{(2)})\frac{c_1(t_d^{(2)})}{N}  
\exp[-S(t_d^{(2)})|_{b_3=b_1}]
\nonumber \\
& &\hspace{0.5cm}
\times[(x_1+x_2)(1+\zeta^{(*)}_d r^2)-1]h^{(2)}_d(x_i,b_i) 
\biggr\}\;,
\label{md}\\
{\cal M}^{(*)}_f&=& 32 \pi\sqrt{2N}{\cal C}_F\sqrt{r}M_B^2 
\int_0^1 [dx]\int_0^{\infty}b_1 db_1 b_2 db_2
\phi_B(x_1)\phi_{D^{(*)}}(x_2)\phi_\pi(x_3) \nonumber \\
& &\times \biggl\{ \alpha_s(t_f^{(1)})\frac{c_1(t_f^{(1)})}{N}  
\exp[-S(t_f^{(1)})|_{b_3=b_2}]
\nonumber \\
& &\hspace{0.5cm}
\times[x_1(1+\zeta^{(*)}_fr^2)
-\zeta^{(*)}_f x_2r^2-x_3(1-r^2)]h^{(1)}_f(x_i,b_i) 
\nonumber \\
& &\hspace{0.5cm}
- \alpha_s(t_f^{(2)})\frac{c_1(t_f^{(2)})}{N}  
\exp[-S(t_f^{(2)})|_{b_3=b_2}]
\nonumber \\
& &\hspace{0.5cm}
\times[(x_1+x_2)(1+\zeta^{(*)}_fr^2)
-\zeta^{(*)}_fr^2]h^{(2)}_f(x_i,b_i) \biggr\}\;,
\label{mf}
\end{eqnarray}
with the definition $[dx]\equiv dx_1dx_2dx_3$. 
The constants are $\zeta_{b,d,f}=-\zeta^{*}_{b,d,f}=1$.

The functions $h^{(j)}$, $j=1$ and 2, appearing in 
Eqs.~(\ref{mb})-(\ref{mf}), are written as
\begin{eqnarray}
\everymath{\displaystyle}
h^{(j)}_b&=& \left[\theta(b_1-b_2)K_0\left(BM_B
b_1\right)I_0\left(BM_Bb_2\right)\right. \nonumber \\
& &\quad \left.
+\theta(b_2-b_1)K_0\left(BM_B b_2\right)
I_0\left(BM_B b_1\right)\right]\;  \nonumber \\
&  & \times \left( \begin{array}{cc}
 K_{0}(B_{j}M_Bb_{2}) &  \mbox{for $B_{j} \geq 0$}  \\
 \frac{i\pi}{2} H_{0}^{(1)}(|B_{j}|M_Bb_{2})  & \mbox{for $B_{j} \leq 0$}
  \end{array} \right)\;,           
\\
\everymath{\displaystyle}
h^{(j)}_d&=& \left[\theta(b_1-b_2)K_0\left(DM_B
b_1\right)I_0\left(DM_Bb_2\right)\right. \nonumber \\
& &\quad \left.
+\theta(b_2-b_1)K_0\left(DM_B b_2\right)
I_0\left(DM_B b_1\right)\right]\;  \nonumber \\
&  & \times \left( \begin{array}{cc}
 K_{0}(D_{j}M_Bb_{2}) &  \mbox{for $D_{j} \geq 0$}  \\
 \frac{i\pi}{2} H_{0}^{(1)}(|D_{j}|M_Bb_{2})  & \mbox{for $D_{j} \leq 0$}
  \end{array} \right)\;,           
\label{hjd}\\
\everymath{\displaystyle}
h^{(j)}_f&=& i\frac{\pi}{2}
\left[\theta(b_1-b_2)H_0^{(1)}\left(FM_B
b_1\right)J_0\left(FM_Bb_2\right)\right. \nonumber \\
& &\quad\left.
+\theta(b_2-b_1)H_0^{(1)}\left(FM_B b_2\right)
J_0\left(FM_B b_1\right)\right]\;  \nonumber \\
&  & \times \left( \begin{array}{cc}
 K_{0}(F_{j}M_Bb_{1}) &  \mbox{for $F_{j} \geq 0$}  \\
 \frac{i\pi}{2} H_{0}^{(1)}(|F_{j}|M_Bb_{1})  & \mbox{for $F_{j} \leq 0$}
  \end{array} \right)\;,           
\end{eqnarray}
with the variables
\begin{eqnarray}
B^{2}&=&x_{1}x_{2}(1-r^{2})\;,
\nonumber \\
B_{1}^{2}&=&(x_{1}-x_{2})x_{3}(1-r^{2})+x_{1}x_{2}(1+r^{2})\;,
\nonumber \\
B_{2}^{2}&=&x_{1}x_{2}(1+r^{2})-(x_{1}-x_{2})(1-x_{3})(1-r^{2})\;,
\nonumber \\
D^{2}&=&F^2=x_{1}x_{3}(1-r^{2})\;, 
\nonumber \\
D_{1}^{2}&=&F_1^2=(x_{1}-x_{2})x_{3}(1-r^{2})\;,
\nonumber \\
D_{2}^{2}&=&(x_{1}+x_{2})r^{2}-(1-x_{1}-x_{2})x_{3}(1-r^{2})\;, 
\nonumber \\
F_{2}^{2}&=&(x_{1}+x_{2})+(1-x_{1}-x_{2})x_{3}(1-r^{2})\;.
\end{eqnarray}
The scales $t^{(j)}$ are chosen as 
\begin{eqnarray}
t_b^{(j)}&=&{\rm max}(BM_B,|B_j|M_B,1/b_1,1/b_2)\;,
\nonumber \\
t_d^{(j)}&=&{\rm max}(DM_B,|D_j|M_B,1/b_1,1/b_2)\; 
\nonumber \\
t_f^{(j)}&=&{\rm max}(FM_B,|F_j|M_B,1/b_1,1/b_2)\;. 
\end{eqnarray}
Here we include also the gluon energies $BM_B$, $DM_B$, and $FM_B$ except 
for the energies $|B_j|M_B$, $|D_j|M_B$, and $|F_j|M_B$ of the internal 
quarks, because the former may not be smaller than the latter.

The corresponding form factors and the nonfactorizable amplitudes for the 
$B\to D^{(*)}\rho$ decays can be computed in a similar way, and their 
expressions are presented in the Appendix B. The only differences are the 
matrix structures of the $\rho$ meson in the calculation of the hard decay 
subamplitudes, and the extra transverse-mode contributions from the $\rho_T$ 
meson, except for the longitudinal-mode contributions from the $\rho_L$ 
meson. The matrix structures associated with the final $\rho_L$ and $\rho_T$ 
mesons are $\not P_3/\sqrt{2N}$ and $\not\epsilon\not P_3/\sqrt{2N}$ with 
$\epsilon\cdot P_3=0$, respectively. We also assume a vanishing $\rho$ meson 
mass. 

In the evaluation of the various form factors and amplitudes, we adopt
$G_F=1.16639\times 10^{-5}$ GeV$^{-2}$, the decay constants $f_B=200$ MeV, 
$f_D=f_{D^*}=220$ MeV, $f_\pi=132$ MeV \cite{A}, and $f_\rho=200$ MeV, the 
CKM matrix elements $|V_{cb}|=0.043$ \cite{L1,WYL1} and $|V_{ud}|=0.974$, 
the masses $M_B=5.28$ GeV, $M_D=1.87$ GeV, and $M_{D^*}=2.01$ GeV 
\cite{PDG}, and the ${\bar B}^0$ ($B^-$) meson lifetime $\tau_{B^0}=1.53$ 
($\tau_{B^-}=1.68$) ps \cite{B}. As to the wave functions, we employ the 
model
\begin{equation}
\phi_{B,D^{(*)}}(x)=\frac{N_{B,D^{(*)}}}{16\pi^2}\frac{x(1-x)^2}
{M_{B,D^{(*)}}^2+C_{B,D^{(*)}}(1-x)}\;,
\label{bw}
\end{equation}
for the $B$ and $D^{(*)}$ mesons, and the Chernyak-Zhitnitsky models from 
QCD sum rules \cite{CZ},
\begin{eqnarray}
\phi_\pi(x)&=&\frac{5\sqrt{6}}{2}f_\pi x(1-x)(1-2x)^2\;,
\label{pwf}\\
\phi^L_{\rho}(x)&=&\frac{5\sqrt{6}}{2}f_{\rho}x(1-x)[0.25(1-2x)^2+0.15]
\;,\label{rhol}\\
\phi^T_{\rho}(x)&=&\frac{5\sqrt{6}}{2}f_{\rho}x^2(1-x)^2\;.
\label{rhow}
\end{eqnarray}
for the pion and the $\rho$ meson, respectively.

The normalization constant $N_B$ and the shape parameter $C_B$ are 
determined by two constraints from the relativistic constituent quark model 
\cite{ASY}. They are given by $N_B=604.332$ GeV$^3$ and $C_B=-27.5$ GeV$^2$, 
which correspond to $f_B=200$ MeV listed above. The shape parameters $C_D$ 
and $C_D^*$ are adjusted such that our predictions for the branching ratios 
of the various modes of $B\to D^{(*)}\pi$ fall into the errors of the 
experimental data \cite{A} shown in Table I. We determine $C_D=-3.372$ 
GeV$^2$ and $C_{D^*}=-3.772$ GeV$^2$, and the corresponding normalization 
constants $N_D=92.85$ GeV$^3$ and $N_{D^*}=119.51$ GeV$^3$ from 
$f_D=f_{D^*}=220$ GeV. Results along with those from the naive PQCD 
formalism based on the full Hamiltonian $H$ ({\it ie.} with $c_1=1$ and 
$c_2=0$), are exhibited in Table I. We find that the predictions for $B$ 
meson decays from these two approaches are close to each other. For 
comparision, we quote the BSW results from \cite{A,NRXS} (BSWI), and from 
\cite{CT} (BSWII), in which a modified pole ansatz is employed to make the 
extraction of the ratio $a_2/a_1$ less mode dependent. 

Different kinds of contributions to the decay amplitudes ${\cal M}_i$ in 
Eqs.~(\ref{M1})-(\ref{M4}) associated with the decays $B\to D^{(*)}\pi$ are 
presented in Table II. It is obvious that the nonfactorizable internal
$W$-emission amplitudes ${\cal M}_d^{(*)}$ play an essential role for the 
explanation of the branching ratios of the $B\to D^{(*)}\pi$ decays: In the 
charged $B$ meson decays ${\cal M}_d^{(*)}$ is about 20\% of the 
factorizable external $W$-emission contributions, while in the neutral $B$ 
meson decays only the factorizable external $W$ emissions dominate, and
all other kinds of contributions are small. Hence, the branching ratios of 
the former are 
predicted to be $(1.2)^2\times (\tau_{B^-}/\tau_{B^0})\approx 1.6$ times of 
the latter, which is well consistent with the data. This conclusion
differs from the previous one drawn in \cite{WYL2}, which is based on 
the naive PQCD formalism: The factorizable internal $W$ emissions give 20\% 
of the
factorizable external $W$-emission contributions, and are responsible for 
the ratio of the charged $B$ to neutral $B$ decay rates. Therefore, we
emphasize that though the values in Column I and in Column II of Table I
are close, the relative weights of the various contributions change.
The $W$-exchange contributions are always negligible, which are only about 
5\% of the external $W$-emission amplitudes. If the conventional 
factorization hypothesis for nonleptonic $B$ meson decays is correct, only 
the diagrams in Fig.~4 are considered. However, our analysis has indicated 
that Figs.~4(c) and 4(d) give small contributions. This is the reason the 
naive choice $\mu=M_b$ for the arguments of the Wilson coefficients in the 
BSW model can not match the data.

The results for the branching ratios of the $B\to D^{(*)}\rho$ decays 
listed in Table I are also satisfactory. Note that after fixing the 
$D^{(*)}$ meson wave function from the data of $B\to D^{(*)}\pi$, there is 
no free parameter left in the analysis of the $B\to D^{(*)}\rho$ decays. 
Hence, the consistency of our predictions with the data is nontrivial. The 
scale dependence of the modified PQCD formalism can be tested simply by 
substituting $2t$ for $t$ in the factorization formulas. The predictions 
decrease a bit as shown in Table I. In the conventional approach of 
effective field theory the substitution of $M_b$ by $2M_b$ for the arguments 
of the Wilson coefficients $c_{1,2}$ results in a more than 20\% difference. 
Hence, the scale setting ambiguity is indeed moderated in our approach.

\vskip 2.0cm
\centerline{\bf IV. The $D\to K^{(*)}\pi$ Decays}
\vskip 0.3cm

We have stated that the naive choice of the BSW parameters 
$a_1=c_1(M_c)+c_2(M_c)/N$ and $a_2=c_2(M_c)+c_1(M_c)/N$ can not explain 
the data of charm decays. To do it, the large $N$ ansatz 
$a_1=c_1(M_c)\approx 1.26$ and $a_2=c_2(M_c)\approx -0.51$ must be assumed
\cite{BSW}. However, the same ansatz $a_2=c_2(M_b)\approx 0.11$ does not 
work for bottom decays, because the best fit to the experimental data 
gives $a_2\approx 0.22$ \cite{CT}. We argue that the large $N$ ansatz is the 
consequence of the factorization hypothesis employed in the BSW model. If 
the nonfactorizable contributions along with the evolution of the Wilson 
coefficients are taken into account, such an ansatz is not necessary. In 
this section we apply the three-scale factorization theorem to the decays 
$D\to K^{(*)}\pi$, and explore in details how the contributions from Figs.~4 
and 5 vary, when they are evaluated at different energy scales. In our 
approach the factorizable and nonfactorizable contributions change with the 
characteristic scales $t$ of the decay processes in different ways, such 
that their combination can explain both the bottom and charm data. That is, 
our work provides a unified viewpoint to the exclusive nonleptonic heavy 
meson decays.

Before proceeding with the calculation of the decay rates, we emphasize that 
the applicability of PQCD to $D$ meson decays with $M_D=1.87$ GeV is 
marginal. It has been shown that the PQCD analysis of exclusive processes 
with the Sudakov effects included is reliable for the energy scale above 
2 GeV \cite{L3}. Therefore, we concentrate only on the mechanism of the 
destructive interference involved in charm decays. For this purpose, it is 
enough to consider the ratios of the charged $D$ meson decay rates to the 
neutral $D$ meson decay rates, 
\begin{equation}
R_1=\frac{{\cal B}(D^-\to K^0\pi^-)}{{\cal B}({\bar D}^0\to K^+\pi^-)}\;,
\;\;\;\;
R_2=\frac{{\cal B}(D^-\to K^{*0}\pi^-)}{{\cal B}({\bar D}^0\to K^{*+}\pi^-)}
\;,
\end{equation}
instead of the individual branching ratios. 

The $D\to K^{(*)}\pi$ decays occur through a similar effective Hamiltonian
\begin{eqnarray}
H_{\rm eff}&=&\frac{G_F}{\sqrt 2}V_{cs}V_{ud}^{*}[c_1(\mu)O_1+ 
c_2(\mu)O_2]\;,
\label{effd}
\end{eqnarray}
where the four-fermion operators are $O_1=({\bar d}u)({\bar s}c)$ and 
$O_2=({\bar s}u)({\bar d}c)$. The analysis of the nonleptonic $B$ meson 
decays in the previous section can be copied to the $D$ meson decays 
directly. The expressions of the decay rates $\Gamma_i$ are similar to 
Eq.~(\ref{dr}), but with the subscripts $i=1$, 2, 3, and 4 denoting the 
modes $D^-\to K^0\pi^-$, ${\bar D}^0\to K^+\pi^-$, $D^-\to K^{*0}\pi^-$ and 
${\bar D}^0\to K^{*+}\pi^-$, respectively. At the same time, the CKM matrix 
element $|V_{cs}|=1.01$ is substituted for $|V_{cb}|$, and the meson masses 
$M_D$ and $M_{K^{(*)}}$ for $M_B$ and $M_{D^{(*)}}$, respectively. In all 
the form factors and nonfactorizable amplitudes the kinematic variables of 
the $B$ $(D^{(*)})$ meson are replaced by those of the $D$ $(K^{(*)})$ 
meson. The ${\bar D}^0$ $(D^-)$ meson lifetime is 
$\tau_{D^0}=0.415$ $(\tau_{D^-}=1.05)$ ps \cite{PDG2}. The $D$ meson wave 
function has been determined in the study of the $B$ meson decays. For the 
kaon, we have the masses $M_K=0.497$ GeV and $M_{K^*}=0.892$ GeV, the decay 
constants $f_K=160$ and $f_{K^*}=220$ MeV, and the model wave functions 
derived from QCD sum rules \cite{CZ},
\begin{eqnarray}
\phi_K(x)&=&\frac{\sqrt{6}}{2}f_Kx(1-x)[3.0(1-2x)^2+0.4]\;,
\label{kw0}\\
\phi^L_{K^*}(x)&=&\frac{\sqrt{6}}{2}f_{K^*}x(1-x)[0.5(1-2x)^2+0.9]\;,
\label{kwl}\\
\phi^T_{K^*}(x)&=&\sqrt{6}f_{K^*}x(1-x)[0.7-(1-2x)^2]\;,
\label{kw}
\end{eqnarray}
for the $K$, $K_L^{*}$ and $K_T^*$ mesons, respectively. Note that we take 
$\phi^L_{K^*}$ as the $K^{*}$ meson wave functions throughout the analysis
of the $D$ meson decays for simplicity. Then all the factorization
formulas in Sec. III can be adopted directly without further modification.
Compared to the pion wave funtion $\phi_\pi$ in Eq.~(\ref{pwf}), $\phi_K$'s 
do not possess dips at the middle of the momentum fraction $x$.
$\phi^T_{K^*}$ has a single hump at $x=1/2$, differing from the behavior
of $\phi_K$ and $\phi^L_{K^*}$. 

Because of the smaller $D$ meson mass, the transverse degrees of freedom
are more important in the definitions of the hard scales $t$. Hence, we
choose the maximum of the scales $1/b_i$ for the arguments $t$ of the Wilson 
coefficients. In this case Sudakov suppression is weaker, and thus 
insufficient to diminish the contributions from the region with $t$ close 
to $\Lambda_{\rm QCD}$, where $c_{1,2}$ diverge. To have meaningful 
predictions, a lower bound $t_c=(1+\epsilon)\Lambda_{\rm QCD}$ must be 
introdiced for the variables $t$ in the numerical analysis, where $\epsilon$ 
is a small number. We then have $t=\max(1/b_i,t_c)$. The results of $R_1$ 
and $R_2$ for $\epsilon\approx 0.0002$ are exhibited in Table III, which are 
well consistent with the data. Note that $t_c$ can be regarded as one and 
the only one free parameter in the analysis of the $D$ meson decays. 
Therefore, the simultaneous fit to $R_1$ and $R_2$ is not trivial.

The contributions from different diagrams are listed in Table IV. The 
$W$-exchange contributions to the neutral $D$ meson decays are negligible
as in the neutral $B$ meson decays. It is easy to observe that with $t$ 
running to below $M_c$, the factorizable internal-$W$ emission contributions 
to the charged $D$ meson decays become very negative due to the evolution 
of $a_2$, and overcome the positive nonfactorizable internal $W$-emission 
amplitudes ${\cal M}_d^{(*)}$. Note that the factorizable internal-$W$ 
emission contributions are positive and small in the $B$ meson decays. The 
naive PQCD formalism based on the full Hamiltonian $H$ without considering 
the evolution of the Wilson coefficients \cite{CM,WYL2} can not account for 
this sign change, since the corresponding coefficient $a_2$ is always equal 
to $1/N$. It then predicts that the charged $D$ meson decay rates are larger 
than the neutral ones (the values of $R_1$ and $R_2$ in Column I of Table 
III are greater than unity) as in the $B$ meson case.

\vskip 2.0cm
\centerline{\bf V. The $B\to J/{\psi}K^{(*)}$ Decays}
\vskip 0.3cm

As mentioned in the Introduction, it has been very difficult to explain the 
ratios $R$ and $R_L$ associated with the $B\to J/{\psi}K^{(*)}$ decays
simultaneously, which were defined in Eq.~(\ref{rrl}), in the BSW framework 
based on the factorization hypothesis. We have observed in Secs. III and IV
that the nonfactorizable contributions play an essential role in the decays 
$B\to D^{(*)}\pi(\rho)$ and $D\to K^{(*)}\pi$. Therefore, it is expected 
that the nonfactorizable effects are also important in the decays 
$B\to J/{\psi}K^{(*)}$. 
In fact, it has been suspected that the discrepancy between 
model-dependent BSW predictions and the experimental data \cite{GKP} is 
attributed to the breakdown of the factorization hypothesis \cite{RA}. 

In this section we apply the three-scale factorization theorem to the
$B\to J/{\psi}K^{(*)}$ decays, and show that our predictions for the 
branching ratios of the various decay modes and for $R$ and $R_L$ are in
good agreement with the data. Similarly, the decays $B\to J/\psi K^{(*)}$ 
occur through the effective Hamiltonian,
\begin{eqnarray}
H_{\rm eff}&=&\frac{G_F}{\sqrt 2}V_{cb}V_{cs}^{*}[c_1(\mu)O_1+ 
c_2(\mu)O_2]\;,
\label{effj}
\end{eqnarray}
with the four-fermion operators $O_1=({\bar s}c)({\bar c}b)$ and 
$O_2=({\bar c}c)({\bar s}b)$. The relevant decay rates have the expression
\begin{equation}
\Gamma_i=\frac{1}{128\pi}G_F^2|V_{cb}|^2|V_{cs}|^2m_B^3\frac{(1-r^2)^3}
{r}|{\cal M}_i|^2\;,
\end{equation}
with $r=M_{J/{\psi}}/M_B$, $M_{J/{\psi}}=3.096$ GeV being the $J/{\psi}$ 
meson mass. The subscript $i=1$ denotes the modes $B^-\to J/{\psi}K^-$ and 
${\bar B}^0\to J/{\psi}K^0$, which possess the same factorization formulas, 
and $i=2$ denotes $B^-\to J/{\psi}K^{*-}$ and ${\bar B}^0\to J/{\psi}K^{*0}$. 
Since the decay amplitudes ${\cal M}_i$ contain only the internal 
$W$-emission contributions from Figs.~4(c) and 4(d) as the factorizable 
part, and from Fig.~5(d) as the nonfactorizable part, their expressions  
are given by
\begin{eqnarray}
{\cal M}_1&=&f_{J/{\psi}}\xi^{(J/\psi)}_{\rm int}+{\cal M}_d^{(J/\psi)}\;,
\label{M9}\\
{\cal M}^L_2&=&f_{J/{\psi}}\xi^{(J/\psi)L}_{\rm int}
+{\cal M}_d^{(J/\psi)L}\;,
\label{M10}\\
{\cal M}^T_2&=&f_{J/{\psi}}\xi^{(J/\psi)T}_{\rm int}
+{\cal M}_d^{(J/\psi)T}\;,
\label{M11}
\end{eqnarray}
where the superscripts $L$ and $T$ denote the logitudinal and transverse 
modes, $B\to J/{\psi}K_L^{*}$ and $B\to J/{\psi}K_T^{*}$, respectively,
and $f_{J/\psi}=390$ MeV is the $J/\psi$ meson decay constant \cite{CM}.
Eq.~(\ref{M9}) is similar to Eq.~(\ref{M1}) and Eqs.~(\ref{M10}) and 
(\ref{M11}) to Eq.~(\ref{M3}), but with the external $W$-emission 
contributions dropped. The type of the mode ${\bar B}^0\to J/{\psi}K^0$ 
corresponds to that of ${\bar B}^0\to D^0\pi^0$, which was not considered 
in Sec. III. Note that the ${\bar B}^0\to D^0\pi^0$ decay involves not only 
the internal $W$-emission but the $W$-exchange diagrams. 

Employing the matrix structure 
$\not \epsilon (\not P_2+M_{J/\psi})/\sqrt{2N}$ for the final $J/\psi$ 
meson, and the vanishing kaon masses $M_K=M_{K^*}=0$ for simplicity, the 
factorization formulas of the form factors and of the nonfactorizable 
amplitudes are derived straightforwardly. They are written as
\begin{eqnarray}
\xi^{(J/\psi)}_{\rm int}&=&-\xi^{(J/\psi)L}_{\rm int}
\nonumber \\
&=&16\pi{\cal C}_F\sqrt{r}M_B^2\int_0^1 dx_1dx_3\int_0^{\infty}b_1db_1
b_3db_3\phi_B(x_1)\phi^{L}_{K^{(*)}}(x_3)a_2(t_{\rm int})
\nonumber \\
& &\times \alpha_s(t_{\rm int})
\left[(1+x_3(1-r^2))h_{\rm int}(x_1,x_3,b_1,b_3,m_{\rm int})\right.
\nonumber \\
& &\left.-x_1r^2 h_{\rm int}(x_3,x_1,b_3,b_1,m_{\rm int})
\right]\exp[-S_B(t_{\rm int})-S_K(t_{\rm int})],
\label{pint1} \\
\xi^{(J/\psi)T}_{\rm int}&=&32\pi{\cal C}_F\sqrt{r}M_B^2
\int_0^1 dx_1dx_3\int_0^{\infty}b_1db_1b_3db_3
\phi_B(x_1)\phi^T_{K^*}(x_3)a_2(t_{\rm int})
\nonumber \\
& &\times \alpha_s(t_{\rm int})
rh_{\rm int}(x_1,x_3,b_1,b_3,m_{\rm int})
\exp[-S_B(t_{\rm int})-S_K(t_{\rm int})],
\label{pint2}
\\
{\cal M}_d^{(J/\psi)}&=&-{\cal M}_d^{(J/\psi)L}\;,
\nonumber \\
&=&16\pi\sqrt{2N}{\cal C}_F\sqrt{r}M_B^2
\int_0^1 [dx]\int_0^{\infty}b_1db_1b_2db_2
\phi_B(x_1)\phi_{J/\psi}^L(x_2)\phi^{L}_{K^{(*)}}(x_3)
\nonumber \\
& &\times \biggl\{
\alpha_s(t^{(1)}_d)\frac{c_1(t^{(1)}_d)}{N}
\exp[-S(t^{(1)}_d)|_{b_3=b_1}]
\nonumber\\
& &\hspace{0.5cm}\times
[2-r^2-2(x_1+x_2)(1-r^2)]h^{(1)}_d(x_i,b_i)
\nonumber \\
& &\hspace{0.5cm}
+\alpha_s(t^{(2)}_d)\frac{c_1(t^{(2)}_d)}{N}
\exp[-S(t^{(2)}_d)|_{b_3=b_1}]
\nonumber \\
& &\hspace{0.5cm}\times
[4x_1-r^2-2x_2(1+r^2)-2x_3(1-r^{2})]h^{(2)}_d(x_i,b_i)
\biggr\}\;, 
\label{pint4}\\
{\cal M}_d^{(J/\psi)T}&=&32\pi\sqrt{2N}{\cal C}_F\sqrt{r}M_B^2
\int_0^1 [dx]\int_0^{\infty}b_1db_1b_2db_2
\phi_B(x_1)\phi_{J/\psi}^T(x_2)\phi^T_{K^*}(x_3)
\nonumber \\
& &\times \biggl\{
\alpha_s(t^{(1)}_d)\frac{c_1(t^{(1)}_d)}{N}
\exp[-S(t^{(1)}_d)|_{b_3=b_1}]
\nonumber \\
& &\hspace{0.5cm}\times 2r(1-x_1-x_2) h^{(1)}_d(x_i,b_i)
\nonumber \\
& &\hspace{0.5cm}
-\alpha_s(t^{(2)}_d)\frac{c_1(t^{(2)}_d)}{N}
\exp[-S(t^{(2)}_d)|_{b_3=b_1}]
\nonumber \\
& &\hspace{0.5cm}\times 2r(1-x_1+x_2) h^{(2)}_d(x_i,b_i)
\biggr\}\;, 
\label{pint5}
\end{eqnarray}
The total Sudakov exponent for the nonfactorizable amplitudes is given by
$S=S_B+S_{J/\psi}+S_K$, where $S_{J/\psi}$ ($S_K$) has the same expression 
as $S_{D^{(*)}}$ ($S_\pi$) in Eq.~(\ref{wpe}). The hard functions 
$h_{\rm int}$ and $h_d^{(j)}$, $j=1$ and 2, are the same as those appearing 
Eqs.~(\ref{hint}) and (\ref{hjd}), but with the arguments
\begin{eqnarray}
m_{\rm int}& =& M_B^2-M_{J/\psi}^2\;,\\ 
D^{2}& =& x_1x_3(1-r^{2})\;,  \\
D_1^{2}& =& (1-x_2)x_1(1+r^{2})-(3-2x_2-x_2^2)\frac{r^2}{4}
\nonumber \\
& &+(x_1+x_2-1)x_3(1-r^{2})\;,  \\
D_2^{2}& =& x_1x_2(1+r^{2})+(x_1-x_2)x_3(1-r^{2})
+(x_2-\frac{1}{4})r^2\;. 
\end{eqnarray}
The hard scales $t$ are also similar to those in the analysis of 
the $B\to D^{(*)}\pi$ decays but with the insertion of the above arguments.

To evaluate the form factors and the nonfactorizable amplitudes, we need 
the information of the $J/\psi$ meson wave function. Unfortunately, there 
are not yet convincing models for them. However, it should be most possible 
that the two charm quarks in the $J/\psi$ meson carry equal fractional 
momenta. Hence, we assume, for convenience, that the wave function 
$\phi_{J/\psi}^T$ for the $(J/\psi)_T$ meson with transverse polarization 
possesses the same form as $\phi_\rho^T\propto x^2(1-x)^2$ for the $\rho_T$ 
meson in Eq.~(\ref{rhow}), because $\phi_\rho^T$ has a  maximum at $x=1/2$. 
We further assume that $\phi_{J/\psi}^L$ for the $(J/\psi)_L$ meson with 
longitudinal polarization is proportional to $x^n(1-x)^n$ with $n$ a free 
parameter, which will be determined by the data of the decay 
$B\to J/\psi K$. That is, we propose the model wave functions,
\begin{eqnarray}
\phi_{J/\psi}^L(x)& =&N_{J/\psi}x^n(1-x)^n\;,\\
\phi_{J/\psi}^T(x)& =&\frac{5\sqrt{6}}{2}x^2(1-x)^2\;.
\end{eqnarray}
The constant $N_{J/\psi}$ is related to the normalization condition
\begin{eqnarray}
\int_{0}^{1}dx\phi_{J/\psi}^{L}(x)=\frac{f_{J/\psi}}{2\sqrt{6}}\;.
\label{njp}
\end{eqnarray}
We stress that the particular form of the $J/\psi$ meson wave functions
are not important. We have tried other models, such as 
$x(1-x)\exp[-(1-2x)^2]$, and found that it works equally well. The kaon 
wave functions have been shown in Eqs.~(\ref{kw0})-(\ref{kw}).

The experimental data of the branching ratios of the $B\to J/\psi K^{(*)}$ 
decays, and of $R$ and $R_L$ \cite{DJP} are exhibited in 
Table V. We determine the parameter $n=1.25$ from the branching ratio
${\cal B}(B\to J/\psi K)$, and then the normalization $N_{J/\psi}=0.858$ 
GeV$^3$ from Eq.~(\ref{njp}). After fixing the $J/\psi$ meson wave 
functions, we evaluate the branching ratios of the decays 
$B\to J/\psi K^{*}_L$ and $B\to J/\psi K^{*}_T$. Results and the 
corresponding factorizable and nonfactorizable contributions are presented 
in Table V and VI, respectively. Obviously, both the branching ratios of
the various decay modes, and $R$ and $R_L$ are explained successfully. If 
the nonfactorizable amplitudes ${\cal M}_d^{(*)}$ are excluded, we shall 
have $R=1.48$ and $R_L=0.92$, which is too large. It implies that the 
nonfactorizable contributions are indeed essential for the decays 
$B\to J/\psi K^{(*)}$. 

\vskip 2.0cm
\centerline{\bf V. Discussion}
\vskip 0.5cm

In this paper we have developed a modified PQCD formalism for
the study of the exclusive nonleptonic heavy meson decays, which embodies
effective field theory and factorization theorems. It
involves three scales: the $W$ boson mass $M_W$, the characteristic energy 
$t$ of the decay processes, and the transverse extent $b$ of the mesons. 
The evolution of the Wilson coefficients from $M_W$ to $t$ and of the 
Sudakov factor from $t$ to $1/b$ are established to make the factorization
formulas explicitly $\mu$-independent. The factorizable, 
nonfactorizable and nonspectator contributions from the external $W$ 
emissions, the internal $W$ emissions, and the $W$ exchanges are all taken 
into account, and have been evaluated reliably. 
We emphasize again that the Wilson coefficient appears as a convolution 
factor of the factorization formulas in our analysis, instead of a
constant coefficient as in the conventional approach of effective field
theory.

Basically, the two main controversies in the exclusive nonleptonic heavy 
meson decays, {\it ie.} the extraction of $a_{1,2}$ from bottom and charm 
decays, and the simultaneous explanation of $R$ and $R_L$, have been 
resolved by our formalism: The nonfactorizable external $W$-emission 
contributions ${\cal M}_b^{(*)}$ alone, which are 20\% of the factorizable 
one, account for the data of the $B\to D^{(*)}\pi(\rho)$ decays.
The factorizable internal $W$-emission contributions, becoming negative 
enough to overcome ${\cal M}_b^{(*)}$ in the $D$ meson case, successfully 
explain the data of the $D\to K^{(*)}\pi$ decays. That is, the evolution of 
the Wilson coefficients can lead to the necessary constructive 
and destructive interferences involved in bottom and charm decays. 
While it is the nonfactorizable contributions that make trivial 
to account for the ratios $R$ and $R_L$ associated with the decays
$B\to J/\psi K^{(*)}$.

Note that the free parameters contained in our formalism are less than the 
decay modes considered. Hence, the match of the theoretical predictions 
with the experimental data is nontrivial, and
indicates that we may have explored the correct mechanism responsible for 
the nonleptonic heavy meson decays. It is worthwhile to compute other
decay modes, whose consistency with the data will further justify our 
approach. Inclusive nonleptonic heavy meson decays are another important
subject to which  our formalism can be applied. The nonfactorizable soft 
corrections $U$ will give the fine tuning of our predictions. These topics 
will be investigated in separate works.

\vskip 0.5cm
\centerline{\bf Acknowledgement}
\vskip 0.3cm
We thank C.H. Chang, H.Y. Cheng, and B. Tseng for useful discussions.
This work was supported by the National Science Council of R.O.C. under
the Grant No. NSC-86-2112-M-194-007.
\vskip 2.0cm

\appendix
\section{Two-loop expressions of $C_{1,2}(\mu)$}
{\setcounter{equation}{0}
\renewcommand{\theequation}{A\arabic{equation}}

In this appendix we present the expressions of the Wilson coefficients 
$c_{1,2}(\mu)$ to two loops, which are given in terms of 
$c_\pm=c_1\pm c_2$ by \cite{Buras}
\begin{equation}
c_\pm(\mu)=\left[1+\frac{\alpha_s(\mu)}{4\pi}J_\pm\right]
\left[\frac{\alpha_s(M_W)}{\alpha_s(\mu)}\right]^{d_\pm}
\left[1+\frac{\alpha_s(M_W)}{4\pi}(B_\pm-J_\pm)\right]\;,
\end{equation}
with the constants
\begin{eqnarray}
& &J_\pm=\frac{d_\pm}{\beta_0}\beta_1-\frac{\gamma^{(1)}_\pm}{2\beta_0},
\;\;\;\;d_\pm=\frac{\gamma^{(0)}_\pm}{2\beta_0},
\nonumber\\
& &\gamma^{(0)}_\pm=\pm 12\frac{N\mp 1}{2N}\;,
\nonumber\\
& &\gamma^{(1)}_\pm=\frac{N\mp 1}{2N}\left[-21\pm\frac{57}{N}
\mp\frac{19}{3}N\pm\frac{4}{3}n_f-2\beta_0k_\pm\right]\;,
\nonumber\\
& &B_\pm=\frac{N\mp 1}{2N}[\pm 11+k_\pm]\;.
\end{eqnarray}
The scheme dependent parameters $k_\pm$ are
\begin{eqnarray}
k_\pm &=& 0 \;\;\;\;\;\;\;{\rm NDR} 
\nonumber \\
&=&\mp 4. \;\;\;\;{\rm HV}
\end{eqnarray}
The constants $\beta_0$ and $\beta_1$ have been defined in Eq.~(\ref{12}).
In this paper we adopt the NDR scheme. However, we have tested the 
sensitivity of our predictions to these two schemes, and found that the 
scheme dependence can be absorbed into the meson wave functions. Namely, the 
wave functions vary with the scheme such that the predictions almost remain 
the same.

When the scale $\mu$ in $\alpha_s(\mu)$ evolves from above $M_b$ to
below $M_b$, the flavor number $n_f$ changes from 5 to 4. A similar 
change from $n_f=4$ to 3 occurs as $\mu$ evolves from above $M_c$ to 
below $M_c$. To make $\alpha_s$ continuous at these thresholds, 
$\Lambda_{\rm QCD}$ must change accordingly. However, again, 
the dependence on $\Lambda_{\rm QCD}$ can also be absorbed into the
wave functions, such that our predictions are insensitive to whether
the continuity conditions of $\alpha_s$ are implemented or not. Hence, for
simplicity and within the errors of the data, we assign the value  
$\Lambda_{\rm QCD}=0.2$ GeV, and $n_f=4$ for bottom decays and $n_f=3$
for charm decays in the numerical analysis.

\vskip 0.4cm
}

\section{ Factorization of the $B\to D^{(*)}\rho$ Decays}
{\setcounter{equation}{0}
\renewcommand{\theequation}{B\arabic{equation}}
\vskip 0.3cm

The factorization formulas for the $B\to D^{(*)}\rho$ decays can be 
derived straightforwardly. The only differences from the $B\to D^{(*)}\pi$ 
case are the matrix structures of the $\rho$ meson in the calculation of
the hard decay subamplitudes $H$, and the extra contributions from the 
transverse modes involving the $\rho_T$ meson, as stated in Sec. III.

The decay rates are given by Eq.~(\ref{dr}) with $i=1$, 2, 3 and 4 
representing the modes $B^-\to D^0\rho^-$, ${\bar B}^0\to D^+\rho^-$, 
$B^-\to D^{*0}\rho^-$ and ${\bar B}^0\to D^{*+}\rho^-$, respectively. The 
decay amplitudes ${\cal M}_i$ are written as
\begin{eqnarray}
{\cal M}_{1}&=&f_{\rho}[(1+r)\xi_{+}-(1-r)\xi_{-}]
+f_D\xi_{\rm int}-{\cal M}_b+{\cal M}_d\;,
\label{rM1}\\
{\cal M}_2&=&f_{\rho}[(1+r)\xi_{+}-(1-r)\xi_{-}]
+f_B\xi_{\rm exc}-{\cal M}_b+{\cal M}_f\;,
\label{rM2}\\
{\cal M}^L_3&=&\frac{1+r}{2r}f_{\rho}
[(1+r)\xi_{A_1}-(1-r)(r\xi_{A_2}+\xi_{A_3})]
\nonumber \\
& &+f_{D^*}\xi^{*}_{\rm int}+{\cal M}_b^{*}-{\cal M}_d^{*}\;,
\label{rM3}\\
{\cal M}_3^{T}&=&f_{D^*}\xi^{T*}_{\rm int}+{\cal M}_d^{T*}\;,
\label{rtM3}\\
{\cal M}^L_4&=&\frac{1+r}{2r}f_{\rho}
[(1+r)\xi_{A_1}-(1-r)(r\xi_{A_2}+\xi_{A_3})]
\nonumber \\
& &+f_B\xi^{*}_{\rm exc}+{\cal M}_b^{*}-{\cal M}_f^{*}\;,
\label{rM4}\\
{\cal M}_4^{T}&=& f_B\xi^{T*}_{\rm exc}+{\cal M}_f^{T*}\;,
\label{rtM4}
\end{eqnarray}
where the superscript $L$ ($T$) denotes the longitudinal (transverse) mode
$B\to D^*\rho^{L(T)}$. We have used the the same notations as
those for the $B\to D^{(*)}\pi$ decays without confusion.

The form factors $\xi_i$, $i=+$, $-$, $A_1$, $A_2$ and $A_3$, related
only to the $B\to D^{(*)}$ transitions, are the same as those associated 
with the $B\to D^{(*)}\pi$ decays. The form factors $\xi^{(*)}_{\rm int}$ 
and $\xi^{(*)}_{\rm exc}$ and the nonfactorizable amplitudes  
${\cal M}^{(*)}_{b,d,f}$ are similar to those in the $B \to D^{(*)}\pi$ 
decays, but with the pion wave function $\phi_\pi(x_3)$ replaced by the 
$\rho_L$ meson wave function $\phi^L_\rho(x_3)$ given in Eq.~(\ref{rhol}). 
The Sudakov exponents $S_\rho$ for the $\rho$ meson and $S_\pi$ for the 
pion are the same. Below we give only the form factors 
$\xi^{T*}_{\rm int}$ and $\xi^{T*}_{\rm exc}$ and the nonfactorizable 
amplitudes ${\cal M}^{T*}_{d,f}$ involved in the transverse modes 
$B\to D^*\rho_T$,
\begin{eqnarray}
\xi^{T*}_{\rm int} 
&=& 16\pi{\cal C}_F\sqrt{r}M_B^2
\int_0^1 dx_1dx_3\int_0^{\infty}b_1db_1b_3db_3
\phi_B(x_1)\phi^{T}_{\rho}(x_3)a_2(t_{\rm int})
\nonumber \\
& &\times \alpha_s(t_{\rm int})
rh_{\rm int}(x_1,x_3,b_1,b_3,m_{\rm int})
\exp[-S_B(t_{\rm int})-S_\rho(t_{\rm int})]\;,
\label{rtint} \\\
\xi^{T*}_{\rm exc}
&=& 16\pi{\cal C}_F\sqrt{r}M_B^2
\int_0^1 dx_2dx_3\int_0^{\infty}b_2db_2b_3db_3
\phi_{D^{*}}(x_2)\phi^{T}_{\rho}(x_3)a_2(t_{\rm exc})
\nonumber \\
& &\times\alpha_s(t_{\rm exc})
r^2 h_{\rm exc}(x_2,x_3,b_2,b_3,m_{\rm exc})
\exp[-S_{D^*}(t_{\rm exc})-S_\rho(t_{\rm exc})]\;,
\label{rtexc}
\nonumber \\
& &\\
{\cal M}^{T* }_d
&=&32\pi\sqrt{2N}{\cal C}_F\sqrt{r}M_B^2
\int_0^1 [dx]\int_0^{\infty}b_1db_1b_2db_2
\phi_B(x_1)\phi_{D^{*}}(x_2)\phi^{T}_{\rho}(x_3)
\nonumber \\
& &\times \biggl\{
\alpha_s(t^{(1)}_d)\frac{c_1(t^{(1)}_d)}{N}
\exp[-S(t_d^{(1)})|_{b_3=b_1}]r[1-x_1-x_2]h_d^{(1)}(x_i,b_i)
\nonumber \\
& &\hspace{0.5cm}
+ \alpha_s(t^{(2)}_d)\frac{c_1(t^{(2)}_d)}{N}
\exp[-S(t_d^{(2)})|_{b_3=b_1}]r(x_1-x_2) h_d^{(2)}(x_i,b_i)\biggr\}\;,
\label{rtI5}
\nonumber \\
& &\\
{\cal M}^{T*}_f
&=&32\pi\sqrt{2N}{\cal C}_F\sqrt{r}M_B^2
\int_0^1 [dx]\int_0^{\infty}b_1db_1b_2db_2
\phi_B(x_1)\phi_{D^{*}}(x_2)\phi^{T}_{\rho}(x_3)
\nonumber \\
& &\times \biggl\{
\alpha_s(t^{(1)}_f)\frac{c_1(t^{(1)}_f)}{N}
\exp[-S(t^{(1)}_f)|_{b_3=b_2}]r^2[1-x_1-x_2]h_f^{(1)}(x_i,b_i)
\nonumber \\
& &\hspace{0.5cm}
+\alpha_s(t_f^{(2)})\frac{c_1(t^{(2)}_f)}{N}
\exp[-S(t_f^{(2)})|_{b_3=b_2}]r^2(x_1-x_2) h_f^{(2)}(x_i,b_i)\biggr\}\;,
\label{rtI6}
\nonumber \\
\end{eqnarray}
with the Sudakov exponent $S=S_B+S_{D^*}+S_\rho$. The functions 
$h_{\rm int}$, $h_{d}^{(j)}$ and $h_{f}^{(j)}$, $j=1$ and 2, and the hard 
scales $t$ have been defined in Sec. III.

}

\newpage

Table I. Predictions for the branching ratios of the $B\to D^{(*)}\pi(\rho)$ 
decays from the PQCD formalism based on $H$ (I), 
on $H_{\rm{eff}}$ (II), on $H_{\rm{eff}}$ but with the
hard scale $t$ replaced by $2t$ (III), and from the BSW 
model with the parameters $a_1=1.15$ and $a_2=0.26$ (BSWI) \cite{A,NRXS} 
and with $a_1=1.012$ and $a_2/a_1=0.224$ (BSWII) \cite{CT}. The CLEO 
data \cite{A} are also shown. 

\vskip 1.0cm
\[\begin{array}{ccccccc}\hline\hline
{\rm modes}&{\rm I(\%)}&{\rm II(\%)}&{\rm III(\%)}&{\rm BSWI}(\%) 
&{\rm BSWII(\%)}&{\rm data(\%)}\\ 
\hline
B^-\to D^0\pi^-&0.52&0.50 &0.47 &0.57& 0.51&0.534\pm 0.025\\
{\bar B}^0\to D^+\pi^-&0.33&0.33 &0.31 &0.35&0.28&0.308\pm 0.026\\
B^-\to D^{*0}\pi^-&0.49&0.48 &0.46 &0.56&0.56&0.497\pm 0.044\\
{\bar B}^0\to D^{*+}\pi^-&0.32&0.32 &0.30 &0.34&0.27&0.304\pm 0.024\\
B^-\to D^0\rho^-&1.34&1.21 &1.16 &1.07&1.11&1.022\pm 0.067\\
{\bar B}^0\to D^+\rho^-&0.63&0.68 &0.62 &0.82&0.69&0.861\pm 0.078\\
B^-\to D^{*0}\rho^-&1.34&1.62 &1.33 &1.27&1.48&1.444\pm 0.134\\
{\bar B}^0\to D^{*+}\rho^-&0.58&0.83 &0.69 &0.93&0.83&0.844\pm 0.071\\
\hline
\end{array}\]

\newpage
Table II. Contributions to the $B\to D^{(*)}\pi$ decays from the 
factorizable external $W$ emissions and internal $W$ emissions (or $W$ 
exchanges), and from the nonfactorizable amplitudes ${\cal M}^{(*)}_{b,d,f}$ 
in Eqs.~(\ref{M1})-(\ref{M4}). The unit is $10^{-3}$ GeV.

\vskip 1.0cm
\[\begin{array}{ccccc}\hline\hline
{\rm amplitudes}&{\rm external}\; W&{\rm internal}\;W &M_b^{(*)}&M_d^{(*)}\\ 
                &{\rm (factorizable)}&{\rm (factorizable)} & &\\ 
\hline
{\cal M}_1 &95.1&2.5&-5.3+14.8i&18.5-10.4i\\
{\cal M}_3 &86.5&2.6&6.6-20.6i&17.0-10.8i\\
\hline \hline
{\rm amplitudes}&{\rm external}\; W&W\;{\rm exchange} &M_b^{(*)}&M_f^{(*)}\\ 
                &{\rm (factorizable)}&{\rm (factorizable)} &&\\ 
\hline
{\cal M}_2 &95.1&-0.6+0.4i&-5.3+14.8i &2.2-3.1i\\
{\cal M}_4 &86.5&-0.6+1.3i&6.6-20.6i &3.0-3.1i\\
\hline
\end{array}\]

\newpage
Table III. Predictions for the ratios $R_1$ and $R_2$ associated with the 
$D\to K^{(*)}\pi$ decays from the PQCD formalism based on $H$ (I) and on 
$H_{\rm{eff}}$ (II). The experimental data \cite{PDG2} are also shown. 

\vskip 1.0cm
\[\begin{array}{cccc}\hline\hline
{\rm modes}&{\rm I}&{\rm II}&{\rm data}\\ 
\hline
R1&4.96&0.74&0.72\\
R2&5.00&0.35&0.38\\
\hline
\end{array}\]

\newpage
Table IV. Contributions to the $D\to K^{(*)}\pi$ decays from the 
factorizable external $W$ emissions and internal $W$ emissions (or $W$ 
exchanges), and from the nonfactorizable amplitudes 
${\cal M}^{(*)}_{b,d,f}$. The unit is $10^{-3}$ GeV.

\vskip 1.0cm
\[\begin{array}{ccccc}\hline\hline
{\rm amplitudes}&{\rm external}\; W&{\rm internal}\;W &M_b^{(*)}&M_d^{(*)}\\ 
                &{\rm (factorizable)} &{\rm (factorizable)} &&\\ 
\hline
{\cal M}_1 &368.0&-192.4&-19.7+17.1i&18.5-24.4i\\
{\cal M}_3 &752.0&-576.7&44.9-10.3i&119.2-39.8i\\
\hline\hline
{\rm amplitudes}&{\rm external}\; W&W\;{\rm exchange} &M_b^{(*)}&M_f^{(*)}\\ 
                &{\rm (factorizable)} &{\rm (factorizable)} &\\ 
\hline
{\cal M}_2 &368.0&3.6-2.1i&-19.7+17.1i&-10.7+24.7i\\
{\cal M}_4 &752.0&-2.5-37.8i&44.9-10.3i&51.6-5.6i\\
\hline
\end{array}\]

\newpage
Table V. Predictions for the branching ratios of the $B\to J/\psi K^{(*)}$ 
decays from the PQCD formalism based on $H$ (I) and on $H_{\rm{eff}}$ (II), 
and from the BSW model with the parameters $a_1=1.012$ and $a_2/a_1=0.224$ 
(BSWII) \cite{CT}. The CLEO and CDF data \cite{DJP} are also shown. 

\vskip 1.0cm
\[\begin{array}{ccccc}\hline\hline
{\rm modes}&{\rm I(\%)}&{\rm II(\%)}&{\rm BSWII(\%)}&{\rm data(\%)}\\ 
\hline
B^-\to J/\psi K^-&0.11&0.11&&0.102\pm 0.014\\
{\bar B}^0\to J/\psi K^0&0.10&0.10&&0.115\pm 0.023\\
B^-\to J/\psi K^{*-}&0.14&0.15&&0.158\pm 0.047\\
{\bar B}^0\to J/\psi K^{*0}&0.13&0.14&&0.136\pm 0.027\\
R&1.32&1.47&1.84&1.36\pm 0.17\pm 0.04\\
 &    &    &    &1.32\pm 0.23 \pm 0.16({\rm CDF})\\
R_L&0.62&0.56&0.56&0.52\pm 0.07\pm 0.04\\
& & & &0.65\pm0.10\pm0.04({\rm CDF})\\
\hline
\end{array}\]

\newpage
Table VI. Contributions to the $B \to J/\psi K^{(*)}$ decays from the 
factorizable internal $W$ emissions and from the nonfactorizable 
amplitudes ${\cal M}_{d}^{(J/\psi)}$. 
The unit is $10^{-3}$ GeV.

\vskip 1.0cm
\[\begin{array}{ccc}\hline\hline
{\rm amplitudes}&{\rm internal}\;W &M_d^{(J/\psi)}\\ 
                &{\rm (factorizable)} &\\ 
\hline
{\cal M}_1 &126.9&-30.1+4.9i\\
{\cal M}_2^L &121.0&-32.8+0.2i\\
{\cal M}_2^T &36.7&34.8+30.9i\\
\hline
\end{array}\]

\newpage
\centerline{\large \bf Figure Captions}
\vskip 0.5cm

\noindent
{\bf Fig. 1.} (a) Separation of the infrared and hard $O(\alpha_s)$ 
contributions in PQCD. (b) $O(\alpha_s)$ factorization into a soft
function and a hard decay subamplitude. 
\vskip 0.5cm

\noindent
{\bf Fig. 2.} (a) Separation of the hard and harder 
$O(\alpha_s)$ contributions in effective field theory. (b) $O(\alpha_s)$ 
factorization into a ''harder" function and a hard decay subamplitude.
\vskip 0.5cm

\noindent
{\bf Fig. 3.} (a) $O(\alpha_s)$ factorization of a soft function from a
full decay amplitude. (b) $O(\alpha_s)$ three-scale factorization 
formula for a decay amplitude.

\vskip 0.5cm
\noindent
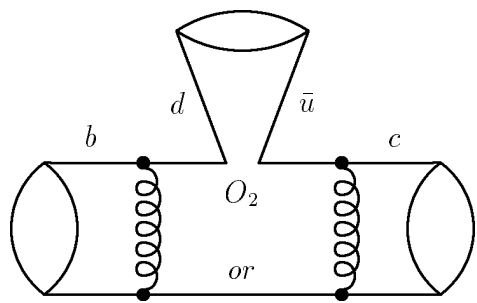
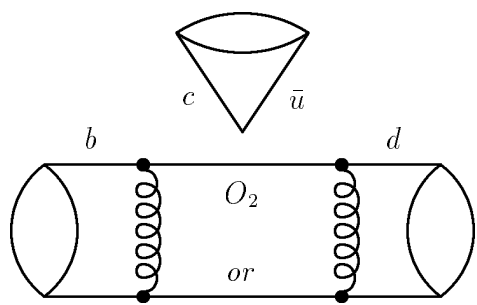
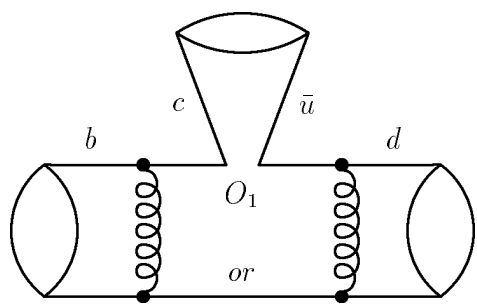
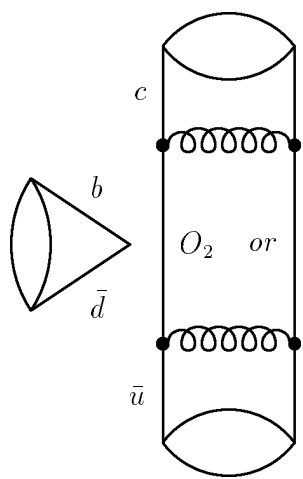
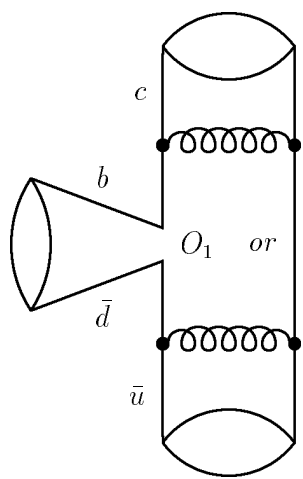
{\bf Fig. 4.} Factorizable external $W$ emissions from (a) the operator 
$O_1$ and from (b) $O_2$, factorizable internal $W$ emissions from (c)
$O_1$ and from (d) $O_2$, and factorizable $W$ exchanges from (e) $O_1$ 
and from (f) $O_2$. 

\vskip 0.5cm
\noindent
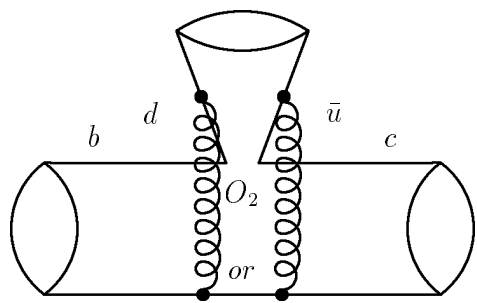
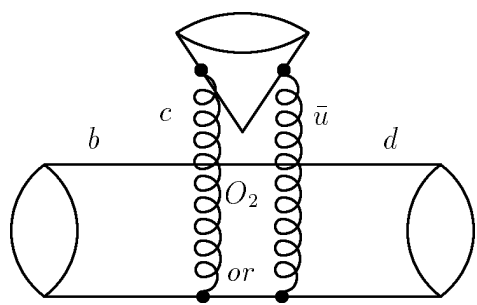
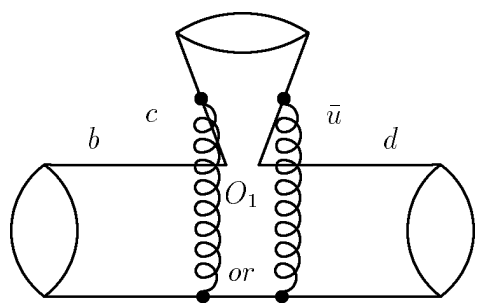
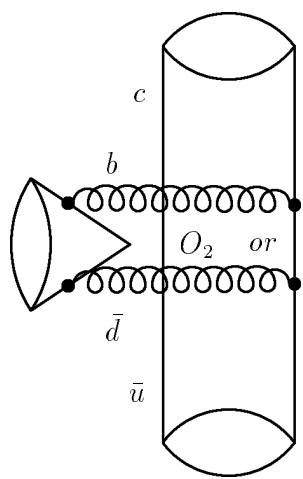
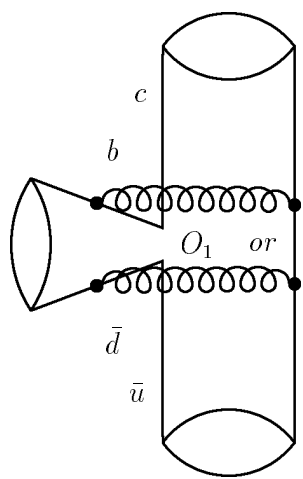
{\bf Fig. 5.} Nonfactorizable external $W$ emissions from (a) the operator 
$O_1$ and from (b) $O_2$, nonfactorizable internal $W$ emissions from (c)
$O_1$ and from (d) $O_2$, and nonfactorizable $W$ exchanges from (e) $O_1$ 
and from (f) $O_2$. 

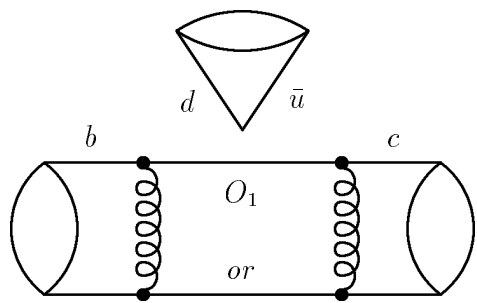
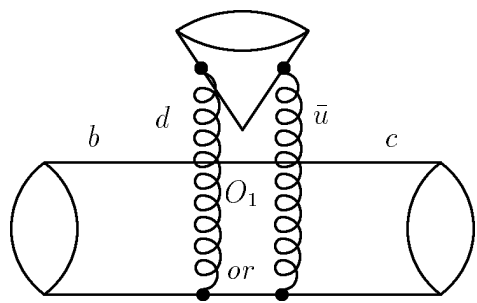
\end{document}